\documentclass[prd,a4paper,superscriptaddress,nofootinbib,10pt]{revtex4} 
\usepackage{amsfonts}
\usepackage{amssymb}
\usepackage{amsmath}
\usepackage{marvosym}
\usepackage{txfonts}
\usepackage[T1]{fontenc}
\usepackage{mathtools}
\usepackage{lscape}
\usepackage{mathrsfs}
\usepackage{graphicx}
\usepackage{bm}
\usepackage{color}
\usepackage{slashed}
\usepackage{hyperref}
\usepackage{tikz}
\usepackage{pgfplots}
\usepackage{array}
\usepackage{textcomp}
\usepackage{epsfig}
\usepackage{slashed}
\usepackage{comment}
\usepackage{xcolor}
\usepackage{enumerate}
\pgfplotsset{width=10cm}

\newcommand{\dd}{\partial}
\newcommand{\ff}{\mathcal{F}}
\newcommand{\ot}{\otimes}
\newcommand{\uu}{\mathcal{U}}
\newcommand{\g}{\mathfrak{g}}
\newcommand{\uf}{\mathcal{U}^\ff(\mathfrak{g})}
\newcommand{\mm}{\mathcal{M}}
\newcommand{\oo}{\mathring}
\newcommand{\lie}{\pounds}

\newcommand{\vp}{\varphi}

\begin{document}
\begin{flushleft}
\texttt{RBI-ThPhys-2023-38}\\
\texttt{ZTF-EP-23-06}
\end{flushleft}

 \title{Metric perturbations in Noncommutative Gravity} 

\author{Nikola Herceg}
\email{nherceg@irb.hr}
\author{Tajron Juri\'c}
\email{tjuric@irb.hr}
\author{Andjelo Samsarov}
\email{asamsarov@irb.hr}
\affiliation{Rudjer Bo\v{s}kovi\'c Institute, Bijeni\v cka  c.54, HR-10002 Zagreb, Croatia}

\author{Ivica Smoli\'c}
\email{ismolic@phy.hr}
\affiliation{Department of Physics, Faculty of Science, University of Zagreb, 10000 Zagreb, Croatia}
\date{\today}

\begin{abstract}
We use the framework of Hopf algebra and noncommutative differential geometry to build a noncommutative (NC) theory of gravity in a bottom-up approach.
Noncommutativity is introduced via deformed Hopf algebra of diffeomorphisms by means of a Drinfeld twist. 
The final result of the construction is a general formalism for obtaining NC corrections to the classical theory of gravity for a wide class of deformations and  a general background. This also includes   a novel   proposal for noncommutative  Einstein manifold. Moreover,  the general construction is applied to the case of a linearized gravitational perturbation theory
to describe a  NC deformation of  the metric perturbations. 
We  specifically present an example for the Schwarzschild background and axial perturbations, which  gives rise to a generalization of the work by  Regge and Wheeler.
All calculations are performed up to  first order in perturbation of the metric and noncommutativity parameter.
The main result is the noncommutative Regge-Wheeler potential.
	Finally, we comment on some differences in properties between the Regge-Wheeler potential and its  noncommutative counterpart.
\end{abstract}

\maketitle
                                                                                                                                           
\section{Introduction}
In over a hundred years, general relativity has yielded some remarkable predictions that have all culminated in impressive experimental confirmation. Probably the most intriguing among them were the predictions of black holes on the one side and gravitational radiation emitted by the massive accelarated astrophysical objects on the other. Over the time, the fascination with black holes and gravitational waves  has been  ever growing, as it was spurred by them persistently evading direct observation.
The fact  that it took so much time and effort for sufficiently improving the required experimental infrastructure has also added
to their mystery.\\

Recently however, this situation has dramatically changed with two epochal discoveries, the latest observations  of  shadows of M87* and Sgr A* \cite{eht1,eht2}, the 
  supermassive black holes located respectively at the center of the Messier 87 galaxy and our own Milky Way galaxy, made mainly by means of the Event Horizon Telescope (EHT),
and the  first ever direct detection of gravitational waves made in 2015 by the LIGO collaboration \cite{ligo}.
Both discoveries marked a kind of a turning point at which something that was previously supposed to be solely a sheer mathematical fiction
 suddenly turned into a tangible physical reality.
In particular is fascinating that
a bright asymmetric ring surrounding a dark central region, a key feature of the supermassive black hole's images  of M87* and Sgr A*, obtained by the  EHT collaboration, 
are well in the range of expected theoretical predictions for a rapidly rotating Kerr black hole made in the framework of the Einstein's general relativity (GR).
Likewise, an almost perfect agreement between the observed and predicted waveforms  as related to two merging black holes from the LIGO experiment 
undoubtedly further confirmed the accuracy of the Einstein's GR.  It is noteworthy to mention that the latter achievement, related to the LIGO's detection of gravitational waves,  also  led to the  Nobel prize for physics in 2017.\\

 While EHT discoveries  provided a novel powerful tool to test Einstein's GR theory in the extremely strong field regime, the discovery of gravitational waves from the binary black hole merger GW150914 by LIGO \cite{ligo} marked the beginning of a new era in gravitational wave astronomy. In particular, the latter has allowed the astronomers to
 dive deeper into the origins of the cosmos and  reach the depths that lie well beyond the boundaries attainable by the cosmic microwave background radiation (CMBR).
As the ability of the CMBR to probe the earlier stages of the universe closer to the  Big Bang is limited
 by the  transparency of the early  universe to the electromagnetic radiation, the reliance on only this type of cosmic messenger will
 fail to provide any information about the universe before its temperature  dropped below $\sim 3000 K$ and the plasma of protons and electrons combined to form atomic hydrogen,
which amounts to the period as early as  few hundred thousand years after the Big Bang.
Unlike  this,  the gravitational waves  have no such type of limitations and are therefore
able to probe  the primordial universe at the Planck scale.\\


Discovery of gravitational waves has also opened up a whole set of new possibilities in  the area of black hole spectroscopy, a term first introduced by Dreyer  et.al. \cite{dreyer},
but the idea that lies behind this notion  had been explored for decades before.
Black hole spectroscopy  relies on the correspondence between black hole spectra and atomic spectra and it consists of measuring of  the complex gravitational-wave frequencies of
black hole merger remnants  and of quantifying their concordance  with the characteristic frequencies of black holes in a ringdown phase
computed at linear order in black hole perturbation theory.\\ 

A notion that is closely related to black hole spectroscopy is that of quasinormal modes (QNMs), the theory for which has been developed during the 1970s, starting with the works by Vishveshwara and Press \cite{vish, press}. First   Vishveshwara  recognized that the response of the black hole to an external perturbation may be represented by a black hole ringdown, which is in turn characterized by a superposition of exponentially damped oscillations, called QNMs, with generally an infinite set of discrete frequencies and damping times,
\begin{equation}  \label{interexpansion}
	h^{\ell m} (t) = \sum_n A^{\ell m}_n e^{-i (\omega_{\ell  m  n} t \: + \: \phi_{\ell m n})},
\end{equation}
where $\omega_{lmn} $ are the quasinormal mode frequencies for the multipole  $(\ell,m),$ and with the overtone number $n$ running unlimitedly from  $n=0$ for the fundamental mode to larger integers labelling higher overtones. Next he realized that the black hole scattering problem is not self-adjoint and that the black hole absorbs gravitational radiation at the horizon and gives off the same at spatial infinity, which has resulted with an appropriate specification of  boundary conditions and complex valued spectrum \cite{vish}. 
 Following that, Press has shown that QNMs take a decisive part in mechanisms that give rise to gravitational radiation and has demonstrated that  the black hole gets rid of the perturbation and returns back to equilibrium only gradually, yielding a long and nearly sinusoidal wave train of gravitational radiation with multipole index $\ell$ \cite{press}.\\

Notably, the analogy between the black hole spectra and  the atomic spectra is not just a matter of a heuristic guesswork, but has a deeper root in an analogy between black hole perturbation theory and quantum mechanics or,  more precisely, an analogy between the theory of scattering of gravitational waves off a  black hole and scattering theory in quantum mechanics. This  analogy comes from noticing  that the separation of the angular variables in the wave equation, using tensor spherical harmonics
with angular indices  $(\ell, m)$, turns the problem of scattering of gravitational  waves  off a Schwarzschild black hole  into a problem that is formally
equivalent to a Schr\"{o}dinger-like equation with a potential  barrier \cite{Thorne:1980ru}. This  has shown to be the case for the axial perturbations of the Schwarzschild black hole,
as demonstrated in a classic paper by Regge and Wheeler \cite{rw}, and also for the polar perturbations of the same spacetime geometry \cite{zer,zer1}.\\

Chandrasekhar and Detweiler calculated, numerically and after using several approximations, the QNM spectra  for both, axial and polar perturbations of  the Schwarzschild black hole and showed their  isospectrality \cite{chandra&detweiler}. For  given  angular momentum indices $(\ell,m),$  they also  computed higher overtones\footnote{For each pair $(\ell,m)$, one finds an infinite series of QNM frequencies $\omega_{\ell m n}$ labeled by the additional integer number $n$, which goes from $n=0$ for the fundamental mode to infinity. Generally, increasing values of $n$ amount to greater imaginary parts and shorter lifetimes for the corresponding overtones.}.
Afterwards, many more semi-classical methods have been developed, which at even deeper level utilized  correlations  between scattering problems in quantum mechanical and gravitational contexts \cite{mashoon,mashoon1,schutz&will}, including those of high level of accuracy \cite{leaver,nollert}.\\

Considering that  realistic astrophysical   objects are non-static and have some angular momentum, a successful separation of the angular part in the perturbation equations for the Kerr black hole, as demonstrated by Teukolsky \cite{teukolsky,teukolsky1,teukolsky2,teukolsky3}, represented a monumental step forward  in a direction of further development of  black hole spectroscopy, as well as  gravitational wave astronomy. In particular, as pointed out in \cite{detweiler,detweiler1}, the fundamental QNM frequency of a Kerr black hole depends on nothing else but its mass and angular momentum, which has a remarkable consequence that, at least in principle, this relationship may be inverted for a purpose of pin pointing the Kerr black hole parameters from the 
 measured values of the frequency and lifetime (which is inversely proportional to the damping width) of the observed quasinormal  mode.\\

Quasinormal modes completely dominate the intermediate phase of the black hole ringdown, which begins immediately after the rapid initial outburst of radiation and lasts all until QNMs become dominated by a power-law behaviour of the perturbation at late times. As shown by Leaver \cite{leaver-expansion} within linear perturbation theory of black holes, each multipole component of the gravitational wave waveform may in this long intermediate interregnum be represented by the expansion  (\ref{interexpansion}), i.e. as a superposition of QNMs.
In linearized Einstein's theory of gravity the QNM frequencies $\omega_{\ell m n}$ for a Kerr black hole \cite{berti,math}  are  known to a very high accuracy, they are generally  complex and  depend only on the mass and the angular momentum of the remnant black hole left after the merger. They do not depend neither on the details of the initial perturbation, nor the way it is triggered. 
As opposed to that, the amplitudes and phases in the expansion (\ref{interexpansion}) depend on the details of the perturbation.\\

The astrophysical processes like binary black hole merger may generally excite multiple quasinormal modes, including different multipoles $(\ell,m),$ as well as different overtones.
It is therefore of interest in black hole spectroscopy/gravitational wave astronomy to single out which combination of modes will have a prevailing influence in the signal.
Recent understandings acquired on the basis of numerical relativity (NR) simulations
\cite{nr1,nr2,nr3,nr4,nr5} suggest that the   $\ell = |m| = 2$ multipole dominates the gravitational wave signal  from a binary black hole merger and that the higher multipoles play the subdominant role.
However, although knowing a precise QNM frequency  for the mode $(\ell = |m| = 2, ~ n =0),$ which is 
the dominant mode in a black hole binary merger, 
might 
help us fix both the mass and the angular momentum of
the  remnant, in practice a single mode is often insufficient  to get accurate  values for these quantities.
Some proposals  have thus appeared in the literature suggesting that some improvements need to be made in estimating the mass and the spin of the remnant.
They included a mixing/coupling of different multipole components \cite{overtones} and an incorporation  of higher overtones \cite{overtones1}.\\

The idea that   the gravitational wave signal  contains contributions from overtones  had thus been proven correct, at least   in the context of inferring the remnant black hole parameters from the observed gravitational wave spectrum. It has also shown to be advantageous in the efforts to lower a mismatch between the waveforms obtained within the framework of linearized black hole perturbation theory in Einstein's GR and those obtained within numerical relativity (NR) framework \cite{nr1,nr2,nr3,nr4,nr5} by using numerical simulations.  Specifically, it has been shown that adding higher overtones
to the $\ell=m=2$ multipole of the radiation lowers the mismatch with NR simulations at least for the times after the peak of the radiation \cite{giesler}.
However, some researchers have raised the doubts  \cite{nr4,Baibhav:2023clw}
about the physical meaning of extending these fits between the waveforms obtained  in linearized GR and NR to the  peak of the  radiation and beyond  toward the earlier times.
Indeed, as the linear perturbation theory in GR  predicts that the frequencies depend only  on the mass and spin of the final Kerr black hole, it is expected to be a good approximation
only at late times, where the intermediate phase of the black hole ringdown roughly falls and when the remnant of the binary merger is close enough to a stationary Kerr black hole.
This would imply that at times before the peak of the radiation, the spacetime is not well described as a linearly perturbed Kerr black hole, and thus the  linearized black hole perturbation theory would not at all be a valid approximation. In this context, the inclusion of higher overtones in order to fit the waveforms obtained by NR simulations for the times before the peak of the radiation would be a rather futile task.\\

On the other side, it is known that
strong gravity and high energy of the fields in vicinity of the horizon create environment in which effects of quantum gravity could emerge.
At these scales the usual point-like structure of spacetime is expected to be replaced with some noncommutative spacetime \cite{Doplicher:1994zv, Doplicher:1994tu}. This should also have the consequences on
the relaxation dynamics of the black hole during the ringdown phase and should 
leave its trace in the gravitational QNM spectrum.\\

In this work we present a systematic way to calculate noncommutative corrections to the equation of motion for metric perturbations. We generalize the calculations of Regge and Wheeler \cite{rw} to noncommutative spacetime and find noncommutative version of the Regge-Wheeler equation and corrections to the Regge-Wheeler potential,
governing the axial perturbations of the Schwarzschild black hole.
There are several ways to construct a theory of noncommutative differential geometry. 
Our approach is based on the Hopf algebra of deformed diffeomorphisms that encode the symmetries of a noncommutative space. By means of Hopf algebra twist we quantize the radial coordinate which effectively smears the horizon at the level of equations of motion. 
Waves of different frequencies and multipoles experience different potentials, especially in the near-horizon region.\\

Before explaining the details that lead to modified perturbation equations and culminate in a successful rephrasing of the problem in terms of 
the quantum-mechanical scattering  described by the Schr\"odinger-type equation, we shall first outline a construction of the noncommutative (NC) gravity in the bottom-up approach, by closely following  
the guidelines presented in \cite{Aschieri:2005zs, Aschieri:2005yw, Aschieri:2009qh, schenkel}. The NC gravity build-up in a bottom-up approach  lies in the core of our analysis, and in particular we implement  it
to a NC deformation of linearized gravitation perturbation theory, as applied to a nonperturbed background of  the Schwarzshild black hole.
Unlike our construction of NC gravity, some other developments in the field
use a top-down approach which starts with an appropriately introduced action principle \cite{Aschieri:2012in}, \cite{Dimitrijevic:2014iwa}, \cite{DimitrijevicCiric:2016qio}, \cite{Aschieri:2022kzo}. There is also a formulation of NC gravity where the corrections are obtained using the NC tetrad formalism \cite{Chaichian:2007we, Chaichian:2007dr, Calmet:2005qm} and coupling the Einstein equation with a NC energy-momentum tensor \cite{Kobakhidze:2007jn}.
References \cite{Chaichian:2007we, Chaichian:2007dr,Calmet:2005qm,Kobakhidze:2007jn} consider NC corrections to the stationary (non-perturbed) black hole. 
Perturbations of the charged scalar field on the NC Reissner-Nordstr\"{o}m background have been studied in \cite{scalar1, scalar2}.
In contrast, our approach to NC perturbation theory is purely geometric, i.e. there are no additional dynamical bosonic or fermionic fields in the spacetime.\\

The paper is structured as follows.
In Section II we introduce the Hopf-algebraic formalism and define the relevant $\star$-differential-geometric structures. 
This section is minimalistic and it briefly summarizes already known facts in the area of Hopf algebras and noncommutative differential geometry.
Section III presents novelties regarding the applications of the formalism of NC differential geometry to gravitational perturbation theory.
As an example, in Section IV we calculate the NC corrections to the metric perturbations around the Schwarzschild blak hole, generalizing the procedure of Regge and Wheeler. We found the NC analog of Regge-Wheeler equation and corrections to the Regge-Wheeler potential. In section V we give our concluding remarks and in the appendices we present some technical details.

\section{Hopf algebra and NC differential geometry}

The symmetries of a manifold are governed by infinitesimal diffeomorphisms, i.e. the Lie algebra of vector fields. Vector fields can act on any product of tensor fields via Leibniz rule. This fact together with the fact that for every vector $v$ there is an inverse $v_{inv}=-v$ gives rise to an algebraic structure called Hopf algebra. The Hopf algebra framework is also well suited for studying the deformed symmetries of NC spaces. The main idea behind this approach is to use the concept of a Drinfeld twist in order to deform  the usual mappings in differential geometry like the algebra of functions, Lie derivative, connection, covariant derivative, curvature and torsion tensors.\\

Let us first state some general facts about Drinfeld twists and Hopf algebras. This is a  minimal and nontechnical intro, for details refer to \cite{Aschieri:2005zs, Aschieri:2005yw, Aschieri:2009qh, schenkel}. Consider a Lie algebra $\mathfrak{g}$ and its universal enveloping algebra $\mathcal{U}(\mathfrak{g})$, an associative unital algebra. It is also a Hopf algebra with the built-in structural maps: coproduct $\Delta:\mathcal{U}(\mathfrak{g})\longrightarrow\mathcal{U}(\mathfrak{g})\otimes\mathcal{U}(\mathfrak{g})$, counit $\epsilon:\mathcal{U}(\mathfrak{g})\longrightarrow\mathbb{C}$ and antipode $S:\mathcal{U}(\mathfrak{g})\longrightarrow\mathcal{U}(\mathfrak{g})$.\\
	$\mathcal{U}(\mathfrak{g})$ has  a canonical Hopf algebra structure. Namely for the generators $g$ of the Lie algebra $\mathfrak{g}$ we have $\Delta(g)=g \otimes \mathbf{1} +\mathbf{1}\otimes g$,   $\epsilon(g)=0$ and $S(g)=-g$.
For the unit element $\mathbf{1} \in \mathcal{U}(\mathfrak{g})$ we have $\Delta(\mathbf{1}) = \mathbf{1} \otimes \mathbf{1}, ~ \epsilon(\mathbf{1}) = 1$ and $S(\mathbf{1}) = \mathbf{1}$. Then using the homomorphism of the coproduct and antihomomorphism property of the antipode, one extends the co-structures to any arbitrary element of $\mathcal{U}(\mathfrak{g})$. In the bialgebra the coproduct $\Delta$ also has to satisfy the following axiom
\begin{eqnarray}
  (\epsilon \otimes id )  \circ \Delta = (id \otimes \epsilon)
  \circ \Delta = id,
\end{eqnarray}
	The coproduct  satisfies the coassociativity condition,
\begin{eqnarray}
  (\Delta \otimes id )  \circ \Delta = (id \otimes \Delta)
	\circ \Delta.
\end{eqnarray}
Antipode $S$ satisfies the axiom
\begin{eqnarray}
    \mu \circ  (S \otimes id )  \circ \Delta =  \mu \circ (id \otimes S)
  \circ \Delta = \eta \circ \epsilon,
\end{eqnarray}
where $\eta$ is a unital map which assigns the unit element $\mathbf{1}$ in the
${\mathcal{U}}(\mathfrak{g})$ to real number $1$.\\ 

As a next step, we take an invertible element $\ff$  in ${\mathcal{U}}(\mathfrak{g}) \otimes   {\mathcal{U}}(\mathfrak{g}) $ and generate a deformation  by means of a series of  similarity transformations implemented
on the structural maps of the Hopf algebra ${\mathcal{U}}(\mathfrak{g})$ thereby turning it into $\uf$.  For any $h \in \mathcal{U}(\mathfrak{g})$ , we carry out the transformation on the 
coproduct, antipode and counit of the original Hopf algebra:
\begin{align}    \label{twistdeformation}
	\Delta^\ff(h) &= \ff \Delta(h) \ff^{-1}, \nonumber  \\
	S^\ff(h) &= \chi S(h) \chi^{-1}, \\
	\epsilon^\ff(h) &=  \epsilon(h),     \nonumber
\end{align}
where $\chi = \mu \circ \big[ (S \ot id) \ff^{-1} \big]$.
The above set of transformations is usually referred to as a deformation by a twist operator or simply twisting.
Upon twisting by $\ff,$ we induce a new  algebraic structure which we denote by $\uf$. As vector spaces, $\uu(\g)$ and $\uf$ are isomorphic. At this point we recall the central result of the deformation theory for Hopf algebras which states that the new structure $\uf$ obtained by the series of deformations (\ref{twistdeformation}) will again constitute a Hopf algebra, but only if the twist element  $\ff$   satisfies the following two conditions
\begin{eqnarray} \label{coccoucon}
	({\mathcal{F}} \otimes {\bf{1}})( \triangle \otimes id){\mathcal{F}} & =&
	({\bf{1}} \otimes {\mathcal{F}})( id \otimes \triangle ){\mathcal{F}}, \nonumber \\
	\mu \circ (\epsilon \otimes id){\mathcal{F}} &= &{\bf{1}} = \mu \circ (id \otimes \epsilon){\mathcal{F}},
\end{eqnarray}
which are known as the $2$-cocycle and counital condition, respectively. The twist operator $\ff$ with the properties (\ref{coccoucon}) is called a Drinfeld twist.
This is an essential result and along with the concept of universal $\mathcal{R}$-matrix and related quasitriangularity properties desribed below,
it has a pivotal role in the construction of NC gravity that we implement in our analysis. Though, before introducing the latter concept, we finish with introducing the elementary 
Hopf algebra setup.\\

Undeformed Hopf algebra has an algebra $\mathcal{A}$ of smooth functions on a manifold $\mm$, $\mathcal{A} = (C^\infty(\mm), \cdot)$, with pointwise multiplication as its module algebra.
The primitivity of its coproduct is reflected as a standard, undeformed Leibniz rule in $C^\infty(\mm),$ since elements of $\uu(\g)$ act as  usual derivations. Twisting of the Hopf algebra $\uu(\g)$ induces the change in its module algebra to accompany for the twisted Leibniz rule. 
As a vector space, the module algebra of $\uf$  is again $C^\infty(\mm)$, but the product changes to the $\star$-product given by:
\begin{equation} \label{starproductdef}
	f \star g = \mu \circ \ff^{-1} (f \ot g) = {\bar{f}}^{\alpha}(f)  {\bar{f}}_{\alpha}(g),
\end{equation}
 for any two functions $f$ and $g$ in the algebra $C^\infty(\mm)$.
Here we introduced a convenient notation for the twist $\ff = f^\alpha \otimes f_\alpha$ and its inverse $\ff^{-1} = \bar{f}^\alpha \otimes \bar{f}_\alpha$, where the sum over $\alpha$ is understood.  From now on,
 $\uf$-module algebra will be denoted by $\mathcal{A}_\star$.\\

As already indicated, another related and  important concept involves a notion of the quasitriangular Hopf algebra. In our context,
the quasitriangular Hopf algebra of interest is the pair  $(\uf,\mathcal{R})$,  consisting of the Hopf algebra $\uf$ and an invertible element $\mathcal{R}= R^{\alpha} \otimes R_{\alpha} \in \uf \otimes \uf,$   satisfying the following three conditions:
\begin{eqnarray} \label{quasitriangularity}
  \Delta^{op}_{\mathcal{F}} (h) &=& \mathcal{R}   \Delta_{\mathcal{F}} (h) \mathcal{R}^{-1},     \nonumber  \\
     (\Delta_{\mathcal{F}}  \otimes  id) \mathcal {R}  &=& \mathcal{R}_{13} \mathcal{R}_{23},    \\
     (id  \otimes \Delta_{\mathcal{F}}) \mathcal{R}  &=& \mathcal{R}_{13} \mathcal{R}_{12},   \nonumber
\end{eqnarray}
for any  $h \in \uf. $ 
Here $ \Delta^{op}_{\mathcal{F}} = \sigma \circ  \Delta_{\mathcal{F}}, ~ {\mathcal{F}}_{op}    = \sigma \circ {\mathcal{F}}$  and $\sigma$ is the flip operator defined as $\sigma(a \ot b) = b \ot a$. In addition, $\mathcal{R}_{13} = R^{\alpha} \otimes \mathbf{1} \otimes  R_{\alpha}$ and similarly for other objects of that kind.  It should be noted that the first of the above conditions also implies that  $  \mathcal{R}= {\mathcal{F}}_{op} {\mathcal{F}}^{-1}.$
Therefore, it is evident  that unlike the Hopf algebra $\uu(\g)$, the twisted Hopf algebra $\uf$ possesses a nontrivial $\mathcal{R}$-matrix given by
\begin{equation}
	\mathcal{R} = \ff_{op}\ff^{-1} = R^\alpha  \otimes  R_\alpha, \quad \mathcal{R}^{-1} = \ff \ff^{-1}_{op} = \bar{R}^\alpha   \otimes  \bar{R}_\alpha.
\end{equation}
On the Hopf algebra level  we say that $\uf$ is quasitriangular\footnote{Even stronger claim holds - it is triangular, meaning $\mathcal{R} \mathcal{R}_{21} = 1 \ot 1$.}.  As the coproducts in $\uf$
 satisfy  $\sigma \circ \Delta_{\mathcal{F}}(h) = \mathcal{R} \Delta_{\mathcal{F}}(h) \mathcal{R}^{-1}$, the module algebra $\mathcal{A}_\star$ will consequentially be $\mathcal{R}$-commutative:
\begin{equation}
	f \star g = \bar{R}^\alpha(g) \star \bar{R}_\alpha(f).
\end{equation}
As already said, properties (\ref{coccoucon}) and (\ref{quasitriangularity}) lie in the core of  our NC gravity construction. 
They are the foundational blocks upon which the entire construction relies, and from which everything else follows - a point that will be demonstrated shortly.
\\

Having completed an introduction of the algebraic part of the whole setup used in the construction, we now turn to the  geometric part.
First we have the $\star$-Lie derivative. For a generic tensor field $\tau$, it  is given by
\begin{equation}
	\pounds^\star_u (\tau) = \pounds_{\bar{f}^\alpha(u)} \: \bar{f}_\alpha (\tau),
\end{equation}
where $\pounds$ is the ordinary Lie derivative.
Here the expressions like ${\bar{f}}^{\alpha}(\tau)$ need to be understood in a sense of the right action,
\begin{equation}
  {\bar{f}}^{\alpha}(\tau) \equiv {\bar{f}}^{\alpha}
    \triangleright \tau,  \quad  
     {\bar{f}}_{\alpha}(\tau) \equiv {\bar{f}}_{\alpha}
    \triangleright \tau,
\end{equation}
where again the right action  is implemented through a Lie derivative action. If ${\bar{f}}^{\alpha}$ itself is a vector field, then we have
\begin{equation}
   {\bar{f}}^{\alpha}
    \triangleright \tau ~ = ~ {\pounds}_{{\bar{f}}^{\alpha} } (\tau), \quad 
   {\bar{f}}_{\alpha}
    \triangleright \tau ~ = ~ {\pounds}_{{\bar{f}}_{\alpha} } (\tau).
\end{equation}
Now,
\begin{itemize}
\item If $\tau$  is a smooth function, then ${\bar{f}}^{\alpha}$  acts as ordinary linear differential operator
\begin{equation} 
  {\pounds}_{{\bar{f}}^{\alpha} } (\tau)
   ~ = ~ {\bar{f}}^{\alpha} (\tau) = {[{\bar{f}}^{\alpha}]}^{\mu} \partial_{\mu} \tau, 
\end{equation}
   i.e. acts as usual derivation.
\item  
    If  $\tau$  is a vector field, then
  \begin{equation}
  {\pounds}_{{\bar{f}}^{\alpha} } (\tau)
   ~ = ~ [{\bar{f}}^{\alpha}, \tau], 
\end{equation}
where  $[~,~]$ is the usual Lie bracket.
\item
   If  $\tau$  is a $1$-form, then one uses Cartan identity
   \begin{equation}
     {\pounds}_{{\bar{f}}^{\alpha} } (\tau)
   ~ = ~ \left( d \circ i_{{\bar{f}}^{\alpha} }  +
    i_{{\bar{f}}^{\alpha} } \circ d \right) \tau,
 \end{equation}
where $~ d ~$  and $~i~$ are the exterior and interior derivatives, respectively.
\item
If  $\tau$  is a general tensor, then one has to use the Leibniz rule for Lie derivative to reduce the expression to Lie derivatives of  vector fields and $1$-forms:
 \begin{equation} 
  {\pounds}_{{\bar{f}}^{\alpha} } (\tau)
   ~ = ~ {\pounds}_{{\bar{f}}^{\alpha} } (\tau_{1}  \otimes  \tau_{2} ) = {\pounds}_{{\bar{f}}^{\alpha} } (\tau_{1}) \otimes  \tau_{2}
  + \tau_{1} \otimes {\pounds}_{{\bar{f}}^{\alpha} } (\tau_{2}).
\end{equation}
\end{itemize}

If on the other hand ${\bar{f}}^{\alpha}$ is an element of the universal enveloping algebra (that is, a
  composition of vector fields), then one uses the usual property of modules realized through a chain of vector fields composed one after another,
 \begin{equation}  
\begin{split}
  {\bar{f}}^{\alpha}
    \triangleright \tau 
   ~ &= ~ X^{\alpha}_1 X^{\alpha}_2 \cdot \cdot \cdot
   X^{\alpha}_k \triangleright \tau  =
    X^{\alpha}_1 \triangleright \left( X^{\alpha}_2   
   \triangleright  \cdot \cdot \cdot  \triangleright 
   \left( X^{\alpha}_k \triangleright \tau  \right)  \right)  \\
 &= X^{\alpha}_1 \triangleright \left( X^{\alpha}_2   
   \triangleright  \cdot \cdot \cdot  \triangleright 
   {\pounds}_{{X}_{k}^{\alpha} } (\tau)  \right),
\end{split}
\end{equation}  
where $ ~ X_{i}^{\alpha}, ~ i= 1,...,k ~$ are vector fields.\\

One can show that $\star$-Lie derivative satisfies 
\begin{align}
	\pounds^\star_u (f \star g) &= \pounds^\star_u(f) \star g + \bar{R}^\alpha(f) \star \pounds^\star_{\bar{R}_\alpha(u)} (g), \\
	\pounds^\star_{f \star u} (g) &= f \star \pounds^\star_u (g)
\end{align}
and generates a $\star$-Lie algebra of vector fields, i.e. deformed infinitesimal diffeomorphisms, satisfying $\star$-Jacobi identities  \cite{Aschieri:2005zs, Aschieri:2005yw, Aschieri:2009qh, schenkel}.\\

In the subsequent analysis we  introduce the following designations: the vector bundle of vector fields is denoted by ${\mathcal{\chi}}_{\star},$ the bundle of $1$-forms is denoted by $ \Omega_{\star}$
and the vector bundle of arbitrary  tensors of order $(p,q)$ is denoted as ${\mathcal T}^{(p,q)}_{\star}.$
Following a deformation of pointwise multiplication, as outlined in (\ref{starproductdef}),
we may continue with implementing the same characteristic deformation pattern when it comes to more general objects  like vector fields, $1$-forms and tensors.
Hence, the product between functions (zero forms) and vectors/$1$-forms is deformed into
\begin{eqnarray} 
   h  \star  u &=&  {\bar{f}}^{\alpha}(h)  {\bar{f}}_{\alpha}(u),   \\
     h  \star  \omega &=&  {\bar{f}}^{\alpha}(h)  {\bar{f}}_{\alpha}(\omega),
\end{eqnarray}
for any $h \in {\mathcal{A}}_{\star}$, and  any vector field $u \in {\mathcal{\chi}}_{\star}$ and $1$-form $\omega \in {\Omega}_{\star}.$
More generally, for any tensor fields $\tau, \tau' \in {\mathcal T}^{(p,q)}_{\star}$
 \begin{equation} \   
     \tau  \otimes_{\star}  \tau'  = {\bar{f}}^{\alpha}(\tau) \otimes  {\bar{f}}_{\alpha}(\tau').
  \end{equation} 
In particular, the wedge product between  two forms  $\omega, \omega' \in   {\Omega}_{\star}$  is deformed as
 \begin{equation}  
     \omega  \wedge_{\star}  \omega'  = {\bar{f}}^{\alpha}(\omega) \wedge  {\bar{f}}_{\alpha}(\omega').
  \end{equation} 
Analogously, a deformation of the Lie bracket $[,]: {\mathcal{\chi}} \times {\mathcal{\chi}} \longrightarrow {\mathcal{\chi}}$ into  $[,]_{\star}: {\mathcal{\chi}}_{\star} \times {\mathcal{\chi}}_{\star} \longrightarrow {\mathcal{\chi}}_{\star}$ is given by
\begin{equation}   
     [u,v]_{\star} = [{\bar{f}}^{\alpha}(u), {\bar{f}}_{\alpha}(v)].
  \end{equation} 
Finally, the  $\star$-pairing, $~{\langle ~, ~\rangle}_{\star} : {\mathcal{\chi}}_{\star}   ~ \times  ~ {\Omega}_{\star}  \rightarrow  {\mathcal{A}}_{\star}~$ between vector fields and $1$-forms can be introduced as
 \begin{equation} 
    {\langle u , \omega \rangle}_{\star} = 
  \langle {\bar{f}}^{\alpha}(u) , {\bar{f}}_{\alpha}(\omega) \rangle.
\end{equation}
Note that as before, at the level of vector spaces, ${\mathcal{\chi}}_{\star}, \Omega_{\star}, {\mathcal T}^{(p,q)}_{\star}$ are isomorphic to 
${\mathcal{\chi}}, \Omega, {\mathcal T}^{(p,q)}.$\\

As a next step in development of NC gravity in a bottom-up approach, it is of essential importance to realize that 
the  $2$-cocycle condition  (\ref{coccoucon}) and  the quasitriangularity properties   (\ref{quasitriangularity}) of the $\mathcal{R}$-matrix  together imply the following relations involving
the $\star$  pairing:
\begin{eqnarray}  
    {\langle h \star u, \omega \star  \tilde{h} \rangle}_{\star} &=& h \star  {\langle  u, \omega  \rangle}_{\star} \star  \tilde{h},    \quad \quad  h, \tilde{h} \in  {\mathcal A}_{\star}, ~ u \in {\mathcal \chi}_{\star},  ~  \omega \in  \Omega_{\star}. \nonumber \\
   {\langle  u,  h \star \omega  \rangle}_{\star} &=&  {\bar{R}}^{\alpha}(h)                     \star  {\langle {\bar{R}}_{\alpha}(u), \omega  \rangle}_{\star}  \label{starpairingprop}, \\
    {\langle  \omega \otimes_{\star} u,  \tau  \rangle}_{\star} &=&    {\langle  \omega,   {\langle  u,  \tau  \rangle}_{\star}  \rangle}_{\star},  
   \quad \quad    u \in {\mathcal \chi}_{\star},  ~  \omega \in  {\mathcal T}^{(0,p)}_{\star}, ~ \tau \in  {\mathcal T}^{(q,s)}_{\star},  \quad q > p.    \nonumber
\end{eqnarray}

Furthermore, a deformed connection  may be introduced as a linear mapping   $~ \hat{\nabla}:  {\mathcal{\chi}}_{\star}   \longrightarrow  \Omega_{\star}   \otimes_{\star}  {\mathcal{\chi}}_{\star}  ~ $  which  satisfies the (undeformed) 
Leibniz rule for all  $h \in {\mathcal{A}}_{\star}$  and  $v \in {\mathcal{\chi}}_{\star},$ 
\begin{equation}   \label{ncconnection}
  \hat{\nabla} (h \star v) = dh \otimes_{\star} v + h \star \hat{\nabla} v.
\end{equation}
Associated with the connection  $\hat{\nabla},$ one may introduce a covariant derivative $\hat{\nabla}_u$  along the vector field $u,$ for any  $u \in {\mathcal{\chi}}_{\star}.$ For all $v \in {\mathcal{\chi}}_{\star}$,
it is defined by
\begin{equation}  \label{nccovariantderivative}
   \hat{\nabla}_u v = {\langle  u,  \hat{\nabla} v  \rangle}_{\star}.
\end{equation}
  The following set of axioms satisfied by the covariant derivative then emerges automatically,
\begin{equation}   \label{covderaxioms}
\begin{split}
  \hat{\nabla}_{u+v} z &= \hat{\nabla}_{u} z + \hat{\nabla}_{v} z,    \\
 \hat{\nabla}_{g \star u} v &= g \star \hat{\nabla}_{u} v,     \\
 \hat{\nabla}_{u} (g \star v) &= {\pounds}^{\star}_u (g ) \star v + {\bar{R}}^{c} (g) \star    
   \hat{\nabla}_{\bar{R}_{c} (u)} v,   
\end{split}
\end{equation}
for all $~ u, v, z \in {\mathcal{\chi}}_{\star} ~$
and $~ g \in {\mathcal{A}}_{\star}.$  Usually, it is customary to introduce the covariant derivative as a linear map that acts on a double copy of vector fields space $( {\mathcal{\chi}}_{\star}  \otimes   {\mathcal{\chi}}_{\star})$ and acquires values on the space of vector fields $({\mathcal{\chi}}_{\star})$ and in addition obeys the set of axioms (\ref{covderaxioms}). However, it has proven as more fruitful to introduce the connection and the associated covariant derivative in a way described above and  simply take advantage of the  properties (\ref{starpairingprop}) of the $\star$-pairing. In this scenario, the  axioms (\ref{covderaxioms})  don't have  to be postulated!  They follow naturally, as a sheer consequence of the properties (\ref{starpairingprop}) and  definitions (\ref{ncconnection}) and (\ref{nccovariantderivative}) for the connection and associated covariant derivative. Here we see that the whole construction outlined so far relies solely on the symmetry properties laid down by the $2$-cocyclicity of the twist operator and the quasitriangularity of the $\mathcal{R}$-matrix. Indeed, the axioms (\ref{covderaxioms}) follow from
the properties (\ref{starpairingprop}), which are in turn consequence of the properties  (\ref{coccoucon}) and    (\ref{quasitriangularity}). \\

It can be seen that covariant derivative satisfies the same deformed Leibniz rule as the Lie derivative ${\pounds}^{\star}_u$. As
in the undeformed case, we define the covariant derivative on functions to be equal to the Lie derivative, for all 
 $h \in {\mathcal A}^{\star}, $$ ~\hat{\nabla}_u (h) = {\pounds}^{\star}_{u}(h)$. Moreover, the last relation in (\ref{covderaxioms}) can be generalized\footnote{As shown in \cite{Aschieri:2009qh}, the actual relation is more complicated for a general type of Drinfeld twists. However, if the twist is composed of affine Killing vector fields in both slots, the  relation simplifies to (\ref{covderaxiomtensor}). } to arbitrary tensors
  $\tau$ and $\tau'$,
\begin{equation}  \label{covderaxiomtensor}
 \hat{\nabla}_{u} (\tau \otimes_{\star} \tau') = \hat{\nabla}_{u} (\tau) \otimes_{\star} \tau' +  {\bar{R}}^{c} (\tau) \otimes_{\star}   
   \hat{\nabla}_{\bar{R}_{c} (u)} (\tau').
\end{equation}

Consider a local frame   $\{ \partial_ {\mu}\}$ of vector fields  and its  $\star$-dual frame   $\{ d x^{\nu} \}$  of $1$-forms;  they are dual with respect to $\star$ pairing. In this sense, they form the $\star$-dual basis\footnote{Dual basis is in general not the same as in commutative setting, but for our application it is.} ${\langle \dd_\mu , dx^\nu \rangle}_{\star} = \delta_\mu{ }^\nu$. The coefficients of affine connection $\hat{\Gamma}^{\mu}_{~\nu \lambda}$ in noncommutative theory may now be introduced in the   usual way as
\begin{equation}  \label{afcoefv}
   \hat{\nabla}_{\partial_\mu} \partial_{\nu} \equiv \hat{\nabla}_{\mu} \partial_{\nu} = \hat{\Gamma}^{\lambda}_{~\mu \nu} \star \partial_{\lambda}.
\end{equation}
Using (\ref{afcoefv}) along with the properties of $\star$ pairing and utilizing the fact that covariant derivative and Lie derivative coincide when acting on scalars, it is possible to extract 
\begin{equation}  \label{afcoeff}
   \hat{\nabla}_{\partial_\mu} dx^{\nu} \equiv \hat{\nabla}_{\mu} dx^{\nu} =  - \hat{\Gamma}^{\nu}_{~\mu \lambda} \star dx^{\lambda}.
\end{equation}

Having  introduced connection and associated covariant derivative, the build-up of NC gravity has got prepared for the remaining and final stage in the construction.
 The  construction  of deformed Riemann curvature  and Ricci tensor now  becomes straightforward. 
The torsion and Riemann curvature tensors are introduced as
\begin{align}  \label{torsion-riemann}
  \hat{T}(u,v) &=  \hat{\nabla}_{u} v - \hat{\nabla}_{\bar{R}^{c} (v)} {\bar{R}_{c} (u)}
  -  [u, v]_{\star}   \equiv {\langle  u \otimes_{\star} v,  \hat{T}   \rangle}_{\star},  \\
 \hat{R}(u,v,z) &=  \hat{\nabla}_{u}\hat{\nabla}_{v} z
  -  \hat{\nabla}_{\bar{R}^{c} (v)}\hat{\nabla}_{\bar{R}_{c} (u)} z -
  \hat{\nabla}_{[u, v]_{\star}} z   \equiv {\langle  u \otimes_{\star} v \otimes_{\star} z,  \hat{R}   \rangle}_{\star},
\end{align}
for all $~ u, v, z \in {\mathcal{\chi}}_{\star}.$
Tensorial character of these quantities is ensured by checking the multilinearity
\begin{equation}  
   \hat{T}(f \star u, v) = f \star \hat{T}(u,v),   \quad  
    \hat{T}(u, f \star v) = {\bar{R}}^{c} (f) \star  \hat{T}({\bar{R}}_{c} (u), v),
\end{equation}
and similarly for the Riemann curvature tensor.
 The components  $\hat{T}^{\lambda}_{~~\mu \nu}$ 
  and  $\hat{R}^{~~~~\lambda}_{\mu \nu \sigma}$ may be determined as
\begin{equation}   \label{torsion-riemann-coef}
\begin{split}
  \hat{T}^{\lambda}_{~~\mu \nu} &= {\langle dx^{\lambda}, \hat{T}(\partial_{\mu}, \partial_{\nu}) \rangle}_{\star},    \\
   \hat{R}^{~~~~\lambda}_{\mu \nu \sigma} &=  {\langle dx^{\lambda}, \hat{R}(\partial_{\mu}, \partial_{\nu}, \partial_{\sigma}) \rangle}_{\star}.
\end{split}
\end{equation}

As for the $\mathcal{R}$-commutativity of the $\star$-product, one can show that the Riemann tensor and the torsion tensor are $\mathcal{R}$-antisymmetric in the first two slots:
\begin{equation}
	\hat{R}(u,v,z) = -\hat{R}(\bar{R}^c (v),\: \bar{R}_c (u),\: z), \qquad \hat{T}(u,v) = -\hat{T}(\bar{R}^c(v), \:\bar{R}_c (u)).
\end{equation}
Unlike the commutative Riemann tensor, this one isn't $\mathcal{R}$-antisymmetric in the last two indices.
As a consequence, NC Ricci tensor defined as
\begin{equation}\label{NCRicci}
	\hat{R}(u,v) = \langle dx^\alpha, \hat{R}(\dd_\alpha, u, v) \rangle_\star
\end{equation}
is not $\mathcal{R}$-symmetric.\\

Finally, the metric  $g$  is an element of   $ \Omega_{\star}   \otimes_{\star}  \Omega_{\star} $  and may be written as
\begin{equation}  
 g= g_{\mu \nu} \star dx^{\mu} \otimes_{\star} dx^{\nu} = g^a \otimes_{\star} g_a  \in  \Omega_{\star}   \otimes_{\star}  \Omega_{\star},
\end{equation}
where the sum over $a$ is understood.
Analogously, the inverse metric $g^{-1} \equiv g^{\star} $ is an element of  ${\mathcal{\chi}}_{\star} \otimes_{\star}  {\mathcal{\chi}}_{\star}, $
$g^{-1} = g^{-1 ~ b} \otimes g_{b}^{-1 } \in   {\mathcal{\chi}}_{\star} \otimes_{\star}  {\mathcal{\chi}}_{\star} $.
These two are related by
\begin{equation}      \label{metricinversecondition}
\begin{split}
	\langle  \langle v, g \rangle_\star, g^{-1} \rangle_{\star} &= \langle v, g^a \rangle_{\star} \star  \langle  g_a, g^{-1 ~ b} \rangle_{\star} \star  g_{b}^{-1 }  = v      \quad  \quad  \mbox{for all}  \quad   v \in    {\mathcal{\chi}}_{\star},        \\
	\langle  \langle \omega, g^{-1} \rangle_\star, g \rangle_{\star} &=
    \langle \omega,  g^{-1 ~ b} \rangle_\star \star  \langle  g_{b}^{-1 }, g^a \rangle_{\star} \star g_a    = \omega
      \quad  \quad  \mbox{for all}  \quad  \omega \in \Omega_{\star}.  
\end{split}
\end{equation}

\section{NC gravity theory}

In the previous section we outlined the Hopf algebra framework and its corresponding NC differential geometry. In order to go a step further and study  physics, i.e. NC gravity, we  still miss one  important ingredient. Namely, we need an equation of motion for the metric $g$, i.e. the equivalent of Einstein's equation in the NC setup. This would determine the dynamics of NC spacetime. In the commutative theory one can define vacuum solutions called Einstein manifolds by imposing that the Ricci tensor is zero, $R_{\mu\nu}(g)=0$. We are primarily interested in generalizing the notion of Einstein manifolds because later on we will be investigating the NC corrections to the metric perturbations around a vacuum solution, that is around a black hole. 
Whilst it may be tempting to  simply postulate $\hat{R}_{\mu\nu}=0,$ considering that this relation  has a correct commutative limit, it  is however in contradiction with the NC differential geometry point of view\footnote{One can also show perturbatively that this relation leads to a system of  partial differential equations that are overcomplete, which means that they give rise to a trivial solution
 where all NC corrections are zero. In the Appendix B it is proven that the equation (\ref{NCeinsymm}) with the  $\mathcal{R}$-symmetrized version of the NC Ricci tensor  admits non trivial solutions.}.
 Namely, postulating $\hat{R}_{\mu\nu}=0$, also implies $\hat{R}_{\nu\mu}=0,$ forcing the NC Ricci tensor \eqref{NCRicci} to be a symmetric tensor, which is in contradiction with the NC geometric definition \eqref{NCRicci}\footnote{In the commutative case this does not pose a problem since $R_{\mu\nu}$ is a symmetric tensor by construction.}. Therefore, we postulate the NC Einstein manifolds as 
\begin{equation}\label{NCein}
\hat{{\rm R}}_{\mu\nu}(g)=0,
\end{equation}
where $\hat{{\rm R}}_{\mu\nu}$ is the $\mathcal{R}$-symmetrized NC Ricci tensor defined by 
\begin{equation}  \label{NCeinsymm}
\hat{{\rm R}}_{\mu\nu}\equiv\frac{1}{2}\left\langle dx^{\alpha}, \hat{R}(\partial_{\alpha}, \partial_{\mu}, \partial_{\nu})+\hat{R}(\partial_{\alpha}, \bar{R}^{A}(\partial_\nu), \bar{R}_{A}(\partial_\mu)\right\rangle_\star.
\end{equation}
Here we have to emphasize that this proposal is different then the one found in \cite{Aschieri:2005zs, Aschieri:2005yw, Aschieri:2009qh, schenkel}. How can \eqref{NCein}  be obtained from  some conveniently adopted action principle or from a generalized Einstein tensor compatible with a $\star$-derived NC Bianchi identity \cite{Aschieri:2005zs, Aschieri:2005yw} is a matter of further investigation and will be reported elsewhere.\\
For calculating NC corrections in \eqref{NCein} we need a specific choice of the twist $\mathcal{F}$. In this paper we will be interested in the Moyal-type twists of the form
\begin{equation}\label{moyal}
\mathcal{F}=\exp\left(i\Theta^{\alpha\beta}V^{1}_\alpha\otimes V^{2}_{\beta}\right),
\end{equation}
where $V^{i}_{\mu}$ are some vector fields and $\Theta^{\mu\nu}$ is a constant antisymmetric tensor. In addition, we require that these two vector fields  commute, thus $\ff$ being an Abelian twist. We will use the standard vector basis $\left\{\partial_{\mu}\right\}$ and demand that basis vectors commute with $V^{i}_{\mu}$. A basis of vector fields of this kind is referred to as a nice basis \cite{schenkel}. Because of the nice basis and Abelian twist, many formulas that we obtain later on will look like their commutative counterparts with the $\star$-product replacing the usual pointwise product.\\
Another important point that we need  to specify before  studying gravity is a choice of  the NC connection $\hat{\nabla}$. The  NC connection is completely determined by the coefficients $\hat{\Gamma}^{\nu}_{~\mu \lambda}$ in \eqref{afcoefv}. Analogously to the commutative case, we define a unique\footnote{The proof of uniqueness can be found in \cite{Aschieri:2009qh}.} torsion free, $\hat{T}=0$, and metric compatible, $\hat{\nabla}(g)=0$, NC Levi-Civita connection
 \begin{equation}\label{gama} 
  \hat{\Gamma}^{\mu}_{~\nu \rho} \equiv \Omega^{\mu}_{~\nu \rho} = \frac{1}{2} g^{\star \mu \alpha} \star \big( \partial_{\nu} g_{\rho \alpha}  +  \partial_{\rho} g_{\nu \alpha} -  \partial_{\alpha} g_{\nu \rho} \big),  
\end{equation}
where $g^{\star \mu \alpha}$ is the unique $\star$-inverted metric satisfying
\begin{equation} \label{gstar}   
	g^{\star \sigma \rho} \star g_{\rho \nu} = \delta^{\sigma}_{~\nu}  \quad \text{ and } \quad   g_{\nu \rho} \star  g^{\star  \rho \sigma} = \delta_{\nu}^{~\sigma}.
\end{equation}
The components of torsion and curvature tensors \eqref{torsion-riemann-coef} are given by
\begin{equation}\label{nct}   
    \hat{T}^{\mu}_{~\nu \rho} = \Omega^{ \mu}_{~ \nu \rho} -  \Omega^{ \mu}_{~ \rho \nu}, 
\end{equation}    
\begin{equation}\label{ncr}  
  \hat{R}^{~~~~\sigma}_{\mu \nu \rho} = \partial_{\mu} \Omega^{\sigma}_{~\nu \rho} - \partial_{\nu} \Omega^{\sigma}_{~\mu \rho} + \Omega^{\beta}_{~\nu \rho} \star \Omega^{\sigma}_{~\mu \beta} - \Omega^{\beta}_{~\mu \rho}  \star \Omega^{\sigma}_{~\nu \beta}.
\end{equation}
Moreover, for the Moyal-type of deformation (\ref{moyal}),  the $\mathcal{R}$-symmetrization reduces to usual symmetrization and we have
\begin{equation}
	\hat{{\rm R}}_{\mu\nu}=\hat{R}_{(\mu\nu)} \equiv \frac{1}{2} (\hat{R}_{\mu\nu} + \hat{R}_{\nu\mu}).
\end{equation}
 The choice of $V^{i}_{\mu}$ and values of $\Theta^{\mu\nu}$ are a matter of quantum-gravity phenomenology and should, at least in principle, be constrained by experiments. In a current lack of such inputs, we will use certain symmetry arguments in order to find solutions to \eqref{NCein} with the deformation \eqref{moyal}. 
 First we may consider an implementation of a symmetry described by some Killing vectors $K_\alpha$ and demand that the deformation  (\ref{moyal}) also respects it. In other words, we require that the twist carrying deformation  is built from these Killing vectors. This can be achieved by defining a Killing twist
\begin{equation}\label{killing}
\mathcal{F}=\exp\left(i\Theta^{\alpha\beta}K_\alpha\otimes K_{\beta}\right).
\end{equation}
It is straightforward to see that any solution to the  commutative Einstein equation  is also a solution of the NC Einstein equation \eqref{NCein}, due to $\pounds_{K_\alpha}g=0$ and the fact that all NC corrections to \eqref{gama}, \eqref{gstar}, \eqref{nct}, \eqref{ncr} vanish \cite{Aschieri:2009qh}. One comes at the same conclusion under even  less stringent conditions, when a deformation is carried out by a semi-Killing twist \cite{Aschieri:2009qh}
\begin{equation}\label{semikilling}
\mathcal{F}=\exp\left(i\Theta^{\alpha\beta}K_\alpha\otimes V_{\beta}\right).
\end{equation}
Note that in the latter case, only one slot is occupied by the Killing vector field.
This is probably one of the reasons why there was not much progress in finding NC corrections to gravity using this formalism\footnote{Apart from the work in \cite{schenkel}.}.
However, we are interested in metric perturbations $h$ around some fixed background $\mathring{g}$, so that the full metric is given by
\begin{equation}
g=\mathring{g}+h.
\end{equation}
For  this reason, we will not build our twist out of Killing vectors of the full metric $g$, but rather out of the Killing vectors of the background $\mathring{g}$. It is interesting to note that for the Killing twist \eqref{killing}, using $\pounds_{K}\mathring{g}=0$ and $\pounds_{K}h\neq 0,$  all NC corrections to \eqref{gama}, \eqref{gstar}, \eqref{nct}, \eqref{ncr} that are linear in $h$ vanish and the lowest non-vanishing NC corrections appear to be quadratic in the metric perturbation $h$. 
\par
The conclusion that can be drawn here is the following: if 
the full symmetry of the background $\mathring{g}$ is to be respected, then the NC effects are inherently nonlinear in perturbation, and one is forced to study quadratic commutative corrections, which would go beyond the linearized theory for metric perturbations. This could be an interesting line of research, but we leave it for a future work. Here, we will rather focus on a semipseudo-Killing twist of the form
\begin{equation}  \label{pseudo-Killingtwist}
	\ff = \exp{\Big(-i \frac{a}{2} \big(K \ot X - X \ot K \big)\Big)},
\end{equation}
where  $a\in\mathbb{R}$, $X$ is some vector field and $K$ is a Killing field of the background metric. In the next section we will show that given $\lie_{X} \oo{g} \neq 0$\footnote{This condition does not hold for the flat Minkowski space since $\lie_{X} \oo{g} = \lie_{X} \eta = 0$ for any $X$, i.e. flat space NC perturbations are of the second order in $h$.}, this twist gives rise to NC corrections that are linear in $h,$ providing us with a NC formalism for linearized metric perturbations.

\section{Linearized noncommutative gravity on a fixed background}

{The twist \eqref{pseudo-Killingtwist}} is a Drinfeld twist, i.e. satisfies the conditions (\ref{coccoucon}).
Our Lie algebra of diffeomorphisms $\g$ is therefore two-dimensional, being generated by $X$ and $K$.
The $\star$-product according to (\ref{starproductdef}) is
\begin{equation}
	f \star g = fg + i\frac{a}{2}\Big(K(f)X(g) - X(f)K(g)\Big) + O(a^2).
\end{equation}
We use now the twist (\ref{pseudo-Killingtwist}) to study a noncommutative gravitational perturbation theory. 
The full metric $g_{\mu \nu}$ is the sum of the background and perturbation $h_{\mu \nu}$, where $|h_{\mu \nu}| \ll |\oo{g}_{\mu \nu}|$.
\begin{equation}  
    g_{\mu \nu} = {\mathring{g}}_{\mu \nu} + h_{\mu \nu},  \quad  \quad  {\mathring{g}}^{\mu \nu} {\mathring{g}}_{\nu \lambda}  = \delta^{\mu}_{~\lambda}.
\end{equation}
As already said, $K$ is the Killing vector field for the background ${\mathring{g}}_{\mu \nu},$
\begin{equation}  
    K({\mathring{g}}_{\mu \nu}) =  {\pounds}_K ({\mathring{g}}_{\mu \nu}) =0,  \quad   K({\mathring{g}}^{\mu \nu}) =  {\pounds}_K ({\mathring{g}}^{\mu \nu}) =0.
\end{equation}
The $h$ metric is assumed to be small relative to $\oo{g}$ and thus we carry out the calculations up to the first order in $h$. 
When switching on a deformation, which is controlled by the parameter of deformation $a$, we also keep only the first order correction terms in $a$. It will turn out that perturbation and deformation parts come in pairs, and are thus always coupled.\\

Due to relative simplicity of the twist (\ref{pseudo-Killingtwist})  and the nice basis, the conditions   (\ref{metricinversecondition})  for the metric inverse simplify significantly and reduce to \eqref{gstar}. Restricting to the  linear perturbations  $(\sim  O(h))$ and the  leading order in NC deformation $(\sim  O(a))$, the solutions to the  above stated conditions give for the inverse metric
\begin{equation}  
    g^{\star \mu \nu} = {\mathring{g}}^{\mu \nu} -  h^{\mu \nu} + {\tilde{g}}^{\mu \nu} = g^{\mu \nu} -  g^{\mu \rho} g_{\rho \lambda} \wedge  g^{\lambda \nu},
\end{equation}
where the following abbreviation was used
\begin{equation} \label{defwedge}
  f \wedge g = i\frac{a}{2} \Big( K(f) X(g) - X(f)  K(g) \Big).
\end{equation}
The inverse metric of $g_{\mu \nu}$ may also be written in the form
\begin{equation} 
	g^{\star \mu \nu} = \oo{g}^{\mu \nu} - \oo{g}^{\mu \alpha} \star h_{\alpha \beta} \star \oo{g}^{\beta \nu}.
\end{equation}
which manifestly shows that a deformation, when present, is always coupled to  perturbation $h_{\mu \nu}.$
It turns out that such $\star$-inverse metric is not symmetric but Hermitian \cite{schenkel}.\\

From the metric compatibility condition and relations (\ref{torsion-riemann-coef}) it then follows that 
coefficients of  the Levi-Civita  connection  in the nice basis, torsion, and Riemann curvature tensor are  in local coordinates given  by \eqref{gama}, \eqref{nct} and \eqref{ncr}.
Using the wedge notation  defined in (\ref{defwedge}) and prefix $\delta$ which stands for just $h$-linear quantities, connection, torsion and Riemann tensor are
\begin{align}   
\begin{split}
\Omega^{\mu}_{~ \nu \rho} &= {\mathring{\Gamma}}^{ \mu}_{~\nu \rho} + \delta {\Gamma}^{ \mu}_{~ \nu \rho} + \bigg( {\mathring{g}}^{\mu \sigma} \wedge \delta {\Gamma}^{ \lambda}_{~ \nu \rho}  \bigg) {\mathring{g}}_{\sigma \lambda}  -  \bigg( h^{\mu \sigma} \wedge {\mathring{\Gamma}}^{ \lambda}_{~\nu \rho}  \bigg) {\mathring{g}}_{\sigma \lambda}   \\ 
 &\equiv   \Gamma^{\mu}_{~ \nu \rho} +  {\tilde{\Omega}}^{\mu}_{~ \nu \rho},
\end{split} \\
\hat{T}^{\mu}_{~\nu \rho} &= \Omega^{ \mu}_{~ \nu \rho} -  \Omega^{ \mu}_{~ \rho \nu} = 0,    \\
\begin{split}
   \hat{R}^{~~~~\sigma}_{\mu \nu \rho}  &=  {\mathring{R}}^{ ~~~~\sigma}_{ \mu \nu \rho} + \delta  R^{~~~~ \sigma}_{\mu \nu \rho} + \partial_{\mu} {\tilde{\Omega}}^{\sigma}_{~\nu \rho} - \partial_{\nu} {\tilde{\Omega}}^{\sigma}_{~\mu \rho} + {\mathring{\Gamma}}^{ \beta}_{~\nu \rho} \wedge \delta {\Gamma}^{ \sigma}_{~ \mu \beta}
     +  \delta {\Gamma}^{ \beta}_{~ \nu \rho} \wedge {\mathring{\Gamma}}^{ \sigma}_{~\mu \beta}     \\    
&+    {\mathring{\Gamma}}^{ \beta}_{~\nu \rho} {\tilde{\Omega}}^{\sigma}_{~\mu \beta}
   + {\tilde{\Omega}}^{\beta}_{~\nu \rho} {\mathring{\Gamma}}^{ \sigma}_{~\mu \beta} -  {\mathring{\Gamma}}^{ \beta}_{~\mu \rho} \wedge   \delta {\Gamma}^{ \sigma}_{~ \nu \beta}  -   \delta {\Gamma}^{ \beta}_{~ \mu \rho} \wedge {\mathring{\Gamma}}^{ \sigma}_{~\nu \beta} - {\mathring{\Gamma}}^{ \beta}_{~\mu \rho}
{\tilde{\Omega}}^{\sigma}_{~\nu \beta} - {\tilde{\Omega}}^{\beta}_{~\mu \rho}  {\mathring{\Gamma}}^{ \sigma}_{~\nu \beta},    \end{split}
\end{align}
where tilde above the quantity  just designates the $a$-linear part.
This leads to  the corrections within the same order in the vacuum  NC Einstein equations \eqref{NCein}.
After acknowledging  the explicit structure of the twist (\ref{pseudo-Killingtwist}) and (\ref{defwedge}), the components for deformed Riemann tensor take the form
\begin{equation}   
	\begin{split}
  \hat{R}^{~~~~\sigma}_{\mu \nu \rho}  &=  {\mathring{R}}^{ ~~~~\sigma}_{ \mu \nu \rho} + \delta  R^{~~~~ \sigma}_{ \mu \nu \rho}  
+ \partial_{\mu} {\tilde{\Omega}}^{\sigma}_{~\nu \rho} - \partial_{\nu} {\tilde{\Omega}}^{\sigma}_{~\mu \rho} 
   + {\tilde{\Omega}}^{\beta}_{~\nu \rho} {\mathring{\Gamma}}^{ \sigma}_{~\mu \beta}    - {\mathring{\Gamma}}^{ \beta}_{~\mu \rho}  {\tilde{\Omega}}^{\sigma}_{~\nu \beta} - {\tilde{\Omega}}^{\beta}_{~\mu \rho}  {\mathring{\Gamma}}^{ \sigma}_{~\nu \beta}.    \\
 &+ i \frac{a}{2} \Big[  K( \delta {\Gamma}^{ \beta}_{~ \nu \rho}) X({\mathring{\Gamma}}^{ \sigma}_{~\mu \beta})
   - X({\mathring{\Gamma}}^{ \beta}_{~\nu \rho}) K(\delta {\Gamma}^{ \sigma}_{~ \mu \beta})
 -K( \delta {\Gamma}^{ \beta}_{~ \mu \rho}) X({\mathring{\Gamma}}^{ \sigma}_{~\nu \beta})
   + X({\mathring{\Gamma}}^{ \beta}_{~\mu \rho}) K(\delta {\Gamma}^{ \sigma}_{~ \nu \beta}) \Big]. 
\end{split}
\end{equation}
NC Ricci tensor defined in (\ref{NCRicci}) is obtained in the nice basis by  simply identifying $\sigma \equiv \mu$ and summing over $\mu,$
\begin{equation}   
\begin{split}
 \hat{R}_{\nu \rho } = \hat{R}^{~~~~\mu}_{\mu \nu \rho}  &=  {\mathring{R}}^{ ~~~~\mu}_{ \mu \nu \rho} + \delta  R^{~~~~ \mu}_{ \mu \nu \rho}  
+ \partial_{\mu} {\tilde{\Omega}}^{\mu}_{~\nu \rho} - \partial_{\nu} {\tilde{\Omega}}^{\mu}_{~\mu \rho}   +    {\mathring{\Gamma}}^{ \beta}_{~\nu \rho} {\tilde{\Omega}}^{\mu}_{~\mu \beta}
   + {\tilde{\Omega}}^{\beta}_{~\nu \rho} {\mathring{\Gamma}}^{ \mu}_{~\mu \beta}    - {\mathring{\Gamma}}^{ \beta}_{~\mu \rho}  {\tilde{\Omega}}^{\mu}_{~\nu \beta} - {\tilde{\Omega}}^{\beta}_{~\mu \rho}  {\mathring{\Gamma}}^{ \mu}_{~\nu \beta}  \\
 &+ i \frac{a}{2} \Big[  K( \delta {\Gamma}^{ \beta}_{~ \nu \rho}) X({\mathring{\Gamma}}^{ \mu}_{~\mu \beta})
   - X({\mathring{\Gamma}}^{ \beta}_{~\nu \rho}) K(\delta {\Gamma}^{ \mu}_{~ \mu \beta}) 
-  K( \delta {\Gamma}^{ \beta}_{~ \mu \rho}) X({\mathring{\Gamma}}^{ \mu}_{~\nu \beta})
   + X({\mathring{\Gamma}}^{ \beta}_{~\mu \rho}) K(\delta {\Gamma}^{ \mu}_{~ \nu \beta}) \Big],
\end{split}
\end{equation}
where
\begin{equation}  
  {\tilde{\Omega}}^{\mu}_{~\nu \rho} =  i \frac{a}{2} \Big[  K( - h^{\mu  \sigma}) X({\mathring{\Gamma}}^{ \lambda}_{~\nu \rho})
   - X({\mathring{g}}^{  \mu \sigma}) K(\delta {\Gamma}^{ \lambda}_{~ \nu \rho}) \Big] {\mathring{g}}_{\sigma \lambda}.
\end{equation}

As $K$ is a Killing vector for the background metric, $K({\mathring{g}}_{\sigma \lambda}) = K({\mathring{g}}^{\sigma \lambda}) =0,$
and $[K, X]= [K, \partial_{\mu}]=0,$ Ricci tensor may be rewritten in the form
\begin{equation}   
\begin{split}
 \hat{R}_{\nu \rho }   &= {\mathring{R}}_{ \nu \rho} + \delta  R_{ \nu \rho} + i \frac{a}{2} {\pounds}_K \Big[-\partial_{\mu} \Big( h^{\mu  \alpha} X({\mathring{\Gamma}}^{ \lambda}_{~\nu \rho}) {\mathring{g}}_{\alpha \lambda} \Big)  -
   \partial_{\mu} \Big( X({\mathring{g}}^{ \mu \alpha})   \delta {\Gamma}^{ \lambda}_{~ \nu \rho}  ~   {\mathring{g}}_{\alpha \lambda} \Big)  
	+  \partial_{\nu} \Big( h^{\mu  \alpha} X({\mathring{\Gamma}}^{ \lambda}_{~\mu \rho})  {\mathring{g}}_{\alpha \lambda} \Big)  + \partial_{\nu} \Big( X({\mathring{g}}^{ \mu \alpha})   \delta {\Gamma}^{ \lambda}_{~ \mu \rho}  ~   {\mathring{g}}_{\alpha \lambda} \Big)  \\  
	&+ \delta {\Gamma}^{ \beta}_{~ \nu \rho}  X({\mathring{\Gamma}}^{ \mu}_{~\mu \beta}) - 
  X({\mathring{\Gamma}}^{ \beta}_{~\nu \rho})  \delta {\Gamma}^{ \mu}_{~ \mu \beta}
  - \delta {\Gamma}^{ \beta}_{~ \mu \rho}  X({\mathring{\Gamma}}^{ \mu}_{~\nu \beta}) +
  X({\mathring{\Gamma}}^{ \beta}_{~\mu \rho})  \delta {\Gamma}^{ \mu}_{~ \nu \beta}
	- {\mathring{\Gamma}}^{ \beta}_{~\nu \rho} h^{\mu  \alpha} X({\mathring{\Gamma}}^{ \lambda}_{~\mu \beta}) {\mathring{g}}_{\alpha \lambda}  - {\mathring{\Gamma}}^{ \beta}_{~\nu \rho}  X({\mathring{g}}^{ \mu \alpha}) \delta {\Gamma}^{ \lambda}_{~ \mu \beta} ~ {\mathring{g}}_{\alpha \lambda}  -  h^{\beta  \alpha} X({\mathring{\Gamma}}^{ \lambda}_{~\nu \rho}) {\mathring{g}}_{\alpha \lambda} {\mathring{\Gamma}}^{ \mu}_{~\mu \beta} \\  
&-    X({\mathring{g}}^{ \beta \alpha}) \delta {\Gamma}^{ \lambda}_{~ \nu \rho} ~ {\mathring{g}}_{\alpha \lambda} {\mathring{\Gamma}}^{ \mu}_{~\mu \beta} + {\mathring{\Gamma}}^{ \beta}_{~\mu \rho} h^{\mu  \alpha} X({\mathring{\Gamma}}^{ \lambda}_{~\nu \beta}) {\mathring{g}}_{\alpha \lambda} 
 +    {\mathring{\Gamma}}^{ \beta}_{~\mu \rho} X({\mathring{g}}^{ \mu \alpha}) \delta {\Gamma}^{ \lambda}_{~ \nu \beta}
~{\mathring{g}}_{\alpha \lambda} +  h^{\beta  \alpha} X({\mathring{\Gamma}}^{ \lambda}_{~\mu \rho}) {\mathring{g}}_{\alpha \lambda}  {\mathring{\Gamma}}^{ \mu}_{~\nu \beta} +  X({\mathring{g}}^{ \beta \alpha}) \delta {\Gamma}^{ \lambda}_{~ \mu \rho} ~ {\mathring{g}}_{\alpha \lambda} {\mathring{\Gamma}}^{ \mu}_{~\nu \beta}  \Big].
\end{split}
\end{equation}

In the next section we will use this NC linearized formalism for the case of Schwarzschild background and axial gravitational perturbations.

\section{Schwarzschild background and NC corrections to the Regge-Wheeler potential}

In 1953 Regge and Wheeler found an equation of motion for axial metric perturbations around the Schwarzschild singularity \cite{rw}. They showed that the equation of motion can be cast into a Schr\"{o}dinger form with the nowadays celebrated Regge-Wheeler potential. A brief overview of their work is given in appendix \ref{app:rw}. In this section we will generalize their work using the NC formalism and show that the equation of motion for the axial metric perturbations can also be cast into a Schr\"{o}dinger form with the NC Regge-Wheeler potential now playing the central role.\\

The background metric $\oo{g}$ is Schwarzschild,
\begin{equation}
d s^2=-\left(1-\frac{R}{r}\right) c^2 d t^2+\frac{1}{1-R / r} d r^2+r^2\left(d \theta^2+\sin ^2 \theta d \vp^2\right), \qquad
	R \equiv \frac{2 G M}{c^2},
\end{equation}
and the twist \eqref{pseudo-Killingtwist} is built out of the following vector fields
\begin{equation} \label{kxdef}
K = \alpha \dd_t + \beta \dd_\varphi, \qquad X = \dd_r.
\end{equation}
Using \eqref{pseudo-Killingtwist} and \eqref{kxdef} one can easily verify that this twist is Drinfeld, Abelian and the vector fields \eqref{kxdef}  commute with the spherical basis $\left\{\partial_t, \partial_r, \partial_\theta, \partial_\varphi\right\}$. The twist built out of $K$ and $X$ from (\ref{kxdef}) produces the following commutation relations between the coordinates
\begin{align}\label{comrel}
	[t\stackrel{\star}{,} r] &= i a \alpha,\\
	[\varphi \stackrel{\star}{,}r] &= i a \beta.
\end{align}
Notice that depending on the choice of the parameters $\alpha$, $\beta$ and $a,$ one can get various Moyal-like NC spaces or $\kappa$-deformed spaces \cite{C1, C2, L1, L2, L3, L4}  when switching to Cartesian coordinates.\\

For the axial metric perturbation we use an ansatz written in the Regge-Wheeler gauge (\ref{rwgauge})
\begin{equation}
\begin{array}{ll}
	h_{t \theta}=\frac{1}{\sin \theta} \sum\limits_{\ell, m} h_0^{\ell m}(r) \partial_{\varphi} Y_{\ell m}(\theta, \varphi)e^{-i \omega t}, & h_{t \varphi}=-\sin \theta \sum\limits_{\ell, m} h_0^{\ell m}(r) \partial_\theta Y_{\ell m}(\theta, \varphi)e^{-i \omega t}, \\
	h_{r \theta}=\frac{1}{\sin \theta} \sum\limits_{\ell, m} h_1^{\ell m}(r) \partial_{\varphi} Y_{\ell m}(\theta, \varphi) e^{-i \omega t}, & h_{r \varphi}=-\sin \theta \sum\limits_{\ell, m} h_1^{\ell m}(r) \partial_\theta Y_{\ell m}(\theta, \varphi) e^{-i \omega t}.
\end{array}
\end{equation}
Now that we know the form of the complete metric $g_{\mu \nu}$, we may take an advantage of the formulas  obtained in the previous section, starting from  the $\star$-metric inverse and all the way up to the NC Ricci tensor. 
Calculations are performed up to the first order in $h_{\mu \nu}$ and noncommutativity parameter $a$.
As discussed in Section III, postulating the equation $\hat{R}_{\mu \nu}=0$ leads to problems. For that specific case we obtain a set of nonseparable equations that admit only trivial solutions. Therefore we want to solve \eqref{NCein}, i.e.
\begin{equation}
	\hat{{\rm R}}_{\mu\nu}=\hat{R}_{(\mu \nu)} = 0.
\end{equation}
It turns out that with this choice we obtain a system of partial differential equations that are separable. 
The radial part will turn out to be governed by three different functions of $r$.
Before writing down the  components of $\hat{R}_{(\mu \nu)}$, we  introduce the parameter $\lambda$ as an eigenvalue of the Killing field's action on the perturbation:
\begin{align}
	h_{\mu \nu} \propto e^{-i \omega t}e^{i m \varphi} \implies \pounds_K h_{\mu \nu} = i \lambda \: h_{\mu \nu}, \\ 
	\text{   for } \quad K = \alpha \dd_t + \beta \dd_\varphi, \quad
	\lambda = -\alpha \omega + \beta m. \label{alfabeta}
\end{align}
As in the commutative case, out of 10 components of the $\hat{\rm R}_{\mu \nu}$, 3 are identically zero and out of 7 remaining, there are only 3 that differ in the radial part. The full $\hat{\rm R}_{\mu \nu}$ is
\begin{equation}
	\hat{\rm R}_{\mu \nu} = 
	\begingroup
\renewcommand*{\arraystretch}{1.6}
\begin{pmatrix}
	0 & 0 & \frac{\dd_\varphi}{\sin \theta}Y_{\ell m}(\theta, \varphi)\frac{- i}{4 r^4} \hat{M}_1(r) & \sin \theta \: \dd_\theta Y_{\ell m}(\theta, \varphi) \hat{M}_1(r) \\
	0 & 0 & \frac{\dd_\varphi}{\sin \theta}Y_{\ell m}(\theta, \varphi)\frac{- i}{4 r^4 (r - R)^2} \hat{M}_2(r) & \sin \theta \dd_\theta Y_{\ell m}(\theta, \varphi) \frac{1}{4 r^4(r - R)^2} \hat{M}_2(r) \\
	\text{Sym} & \text{Sym} & \frac{(\sin \theta \dd^2_{\theta \varphi} - \cos \theta \dd_\varphi)}{1 - \cos^2 \theta}Y_{\ell m}(\theta, \varphi) \frac{i}{r^2} \hat{M}_3(r) & (\sin \theta \dd^2_{\theta \theta} - \cos \theta \dd_\theta - \frac{1}{\sin \theta} \dd^2_{\varphi \varphi}) Y_{\ell m}(\theta, \varphi) \frac{-1}{2 r^2} \hat{M}_3(r) \\
	\text{Sym} & \text{Sym} & \text{Sym} & (\sin \theta  \dd^2_{\theta \varphi} - \cos \theta \dd_\varphi)Y_{\ell m}(\theta, \varphi) \frac{i}{r^2} \hat{M}_3(r)
\end{pmatrix},
\endgroup
\end{equation}
where
\begin{align}
\begin{split}
	\hat{M}_1(r) =&\: 2r (2R - \ell(\ell + 1)r) h_0 + 4 i r^2 \omega (r - R) h_1 + 2 r^3(r - R)(i \omega h_1' + h_0'') \\
	&\quad+ \lambda a \Big(  (2\ell(\ell + 1)  r + R)h_0 + i r \omega (4 r - 3 R) h_1 + r (4r - 5 R) h_0' + r^2 R ( i\omega h_1' + h_0'')  \Big),
\end{split} \label{ricci2} \\  \nonumber \\
\begin{split}
	\hat{M}_2(r) =&\: 4 i r^4 (r - R)\omega h_0 + 2 r^2 (r - R) \big(r^3 \omega^2 - (r - R)(\ell(\ell + 1) - 2)\big)h_1 - 2 i \omega r^5(r - R) h_0'  \\
	&\: + \lambda a \Big( 2 i r^3 \omega (r - 2R) h_0 + \big((2\ell(\ell + 1)+12) r (r- R)^2 - 9(r - R)^2 R - r^4 R \omega^2\big)h_1  \\
	&\: +i r^4 R \omega h_0' + 2 r (r - R)^3 h_1' \Big),
\end{split} \label{ricci1} \\ \nonumber \\
\begin{split}
	\hat{M}_3(r) =&\: \frac{i r^3 \omega}{r - R}h_0 + R h_1 + r(r - R)h_1' - \lambda a \Big( \frac{i r^2 R \omega}{2(r - R)^2}h_0 - 3 \frac{r - R}{r} h_1 - \frac{1}{2}R h_1'\Big).
\end{split} \label{ricci3}
\end{align}
The vacuum Einstein equation (\ref{NCein}) reduces to three radial equations: $\hat{M}_1 =0, ~ \hat{M}_2 = 0, ~ \hat{M}_3 = 0$. 
These equations represent a system of coupled ordinary differential equations of the second order but for only two functions: $h_0$ and $h_1$. It can be shown, however, that these three components are linearly dependent (see appendix \ref{app:dependence}). Therefore, we are free to use any two of the three equations to solve the system.
We express $h_0$ from equation $\hat{M}_3 = 0$ and plug it into equation $\hat{M}_1 = 0$ to get

\begin{equation}\begin{split}
        &r (r-R) \Big( \ell(\ell + 1) r (R-r) + 2r^2 - 6rR + 5R^2 + \omega^2 r^4 \Big) h_1 + r^2(r - R)^2\Big( (5R - 2r)h_1' + r(r - R)h_1''\Big) \\
         &+  \lambda a \bigg[\Big(\ell(\ell + 1)r(r - R)^2 - 6r^3 + \frac{R}{2}(49
r^2 - 64 r R + 26 R^2 - \omega^2 r^4) \Big) h_1 + r(r - R)^2 \Big( 3 ( r- 2R) h_1' + \frac{1}{2}r R h_1''\Big) \bigg]=0.
\end{split} \label{radial1}
\end{equation}
To further reduce this equation to a Schr\"odinger form, we introduce the NC tortoise coordinate $\hat{r}_*$ and modify the $h_1(r)$ function as
\begin{align}
	\frac{d r}{d \hat{r}_*} &= 1 - \frac{R}{r} + \lambda a \frac{R}{2 r^2} \implies \hat{r}_* = r + R \log \frac{r - R}{R} + \frac{\lambda a}{2} \frac{R}{r - R}, \label{kornjacar}\\
	h_1(r) &= \frac{r^2}{r - R}\Big( 1 + \frac{\lambda a}{2} \Big( \frac{3}{r} - \frac{1}{r - R} + \frac{1}{R} \log \frac{r}{r - R} \Big) \Big)\: W(r). \label{kornjacaw}
\end{align}

For details see appendix \ref{app:schrodinger}.
Equation governing the gravitational perturbations becomes
\begin{equation} \label{master}
	\begin{split}
	&\frac{d^2 W}{d \hat{r}_*^2} + \Big( \omega^2 - V(r) \Big) W = 0, \\
		&V(r) = \frac{(r - R)\big(\ell (\ell + 1)r - 3R\big)}{r^4} + \lambda a \frac{\ell(\ell + 1)(3R - 2r)r + R(5r - 8R)}{2r^5}.
	\end{split}
\end{equation}
The first term in the potential $V(r)$ is the usual Regge-Wheeler potential $V_{RW}$ and the second term is the correction coming from the spacetime noncommutativity.\\

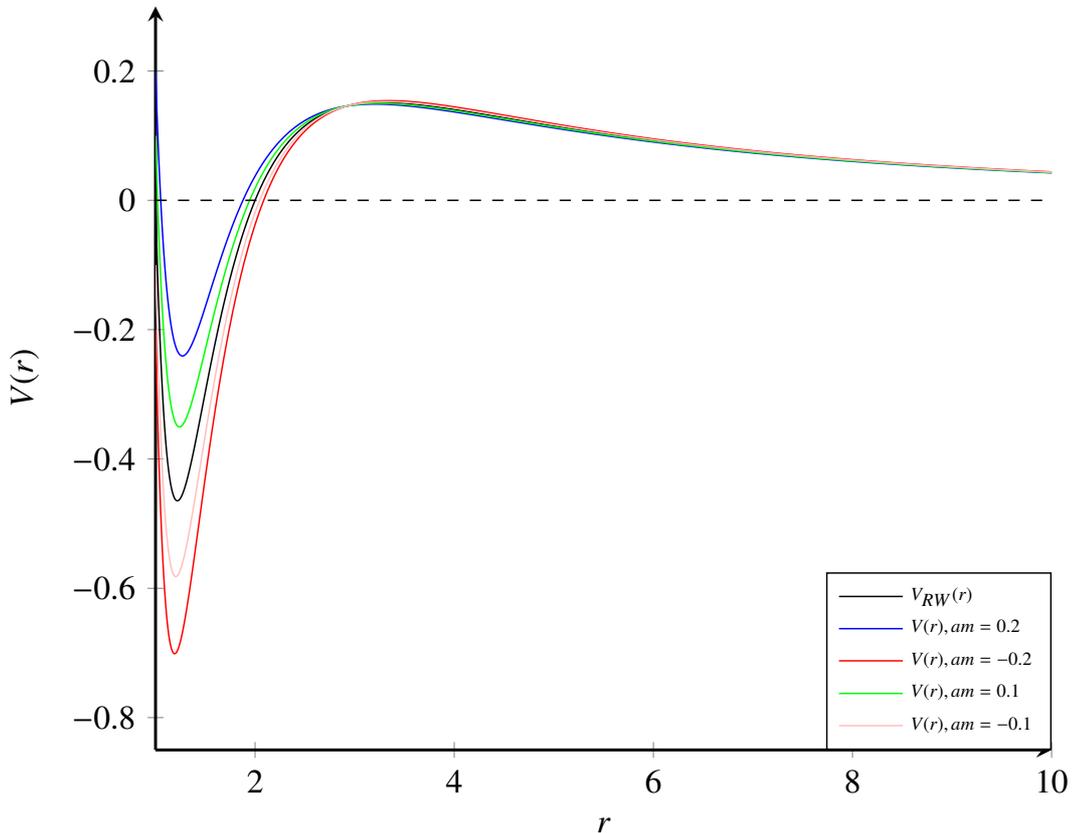
\begin{figure}[h]
\centering
	\begin{tikzpicture}[scale=1.4]
\begin{axis}[
    axis lines = left,
    axis line style = thick,
    xlabel = \(r\),
    ylabel = {\(V(r)\)},
xmin=1,
    xmax=10,
    ymin=-0.85,
    ymax=0.3,
    yticklabel style={/pgf/number format/fixed},
    legend cell align={left},
    legend style={font=\tiny, at={(1,0)},anchor=south east}]
\addplot [
    domain=1:10, 
    samples=900, 
    color=black,
]
{(x-2)*(2*(2+1)*x-6)/x^4};
\addlegendentry{\(V_{RW}(r)\)}

\addplot [
    domain=1:10, 
    samples=900, 
    color=blue,
    ]
    {(x-2)*(2*(2+1)*x-6)/x^4+0.2*(2*(2+1)*(6-2*x)*x+2*(5*x-16))/(2*x^5)};
\addlegendentry{\(V(r), am=0.2\)}

\addplot [
    domain=1:10, 
    samples=900, 
    color=red,
    ]
    {(x-2)*(2*(2+1)*x-6)/x^4-0.2*(2*(2+1)*(6-2*x)*x+2*(5*x-16))/(2*x^5)};
\addlegendentry{\(V(r), am=-0.2\)}

\addplot [
    domain=1:10, 
    samples=900, 
    color=green,
    ]
    {(x-2)*(2*(2+1)*x-6)/x^4+0.1*(2*(2+1)*(6-2*x)*x+2*(5*x-16))/(2*x^5)};
\addlegendentry{\(V(r), am=0.1\)}
    \draw[dashed] (axis cs: 1,0) -- (axis cs: 10,0);

\addplot [
    domain=1:10, 
    samples=900, 
    color=pink,
    ]
    {(x-2)*(2*(2+1)*x-6)/x^4-0.1*(2*(2+1)*(6-2*x)*x+2*(5*x-16))/(2*x^5)};
\addlegendentry{\(V(r), am=-0.1\)}
\end{axis}
\end{tikzpicture}
	\caption{The figure is drawn for $R=2$ ($M=1$) and $\ell=2$ for a wider range of $r$ and $V(r)$.
Around the peak and further away from the horizon, the behaviour appears to be very similar to the commutative case.} \label{grafsve}
\end{figure}

In order to illustrate the NC effects of the potential $V(r),$ we  simplify our problem by choosing $\alpha=0$ and $\beta=1$. In such case $[\varphi \stackrel{\star}{,}r] = i a$ and $\lambda a=am,$ where $m$ is the magnetic quantum number. This is the case of purely angular Killing field $K = \dd_\varphi$, and we see that different multipoles $(\ell, m)$ of the perturbation interact with different gravitational potentials since $\lambda = m$. We plot the potential for the quadrupole  mode $\ell = 2$ and several values of $am=\pm 0.2, \pm 0.1$  in FIG.\ref{grafsve}.

\begin{figure}[h]
\centering
	\begin{tikzpicture}[scale=1.4]
\begin{axis}[
    axis lines = left,
    axis line style = thick,
    xlabel = \(r\),
    ylabel = \(V(r)\),
xmin=1.8,
    xmax=10,
    ymin=0.00,
    ymax=0.16,
    yticklabel style={/pgf/number format/fixed},
    legend cell align={left},
    legend style={font=\tiny,at={(1,0)},anchor=south east}]
\addplot [
    domain=1:10, 
    samples=500, 
    color=black,
]
{(x-2)*(2*(2+1)*x-6)/x^4};
\addlegendentry{\(V_{RW}(r)\)}

\addplot [
    domain=1:10, 
    samples=500, 
    color=blue,
    ]
    {(x-2)*(2*(2+1)*x-6)/x^4+0.2*(2*(2+1)*(6-2*x)*x+2*(5*x-16))/(2*x^5)};
\addlegendentry{\(V(r), am=0.2\)}

\addplot [
    domain=1:10, 
    samples=500, 
    color=red,
    ]
    {(x-2)*(2*(2+1)*x-6)/x^4-0.2*(2*(2+1)*(6-2*x)*x+2*(5*x-16))/(2*x^5)};
\addlegendentry{\(V(r), am=-0.2\)}

\addplot [
    domain=1:10, 
    samples=500, 
    color=green,
    ]
    {(x-2)*(2*(2+1)*x-6)/x^4+0.1*(2*(2+1)*(6-2*x)*x+2*(5*x-16))/(2*x^5)};
\addlegendentry{\(V(r), am=0.1\)}

\addplot [
    domain=1:10, 
    samples=500, 
    color=pink,
    ]
    {(x-2)*(2*(2+1)*x-6)/x^4-0.1*(2*(2+1)*(6-2*x)*x+2*(5*x-16))/(2*x^5)};
\addlegendentry{\(V(r), am=-0.1\)}
\end{axis}
\end{tikzpicture}
\caption{Plot of the potential for $R=2$ ($M=1$) and $\ell=2$, but only around the peak.} \label{grafzoom}
\end{figure}
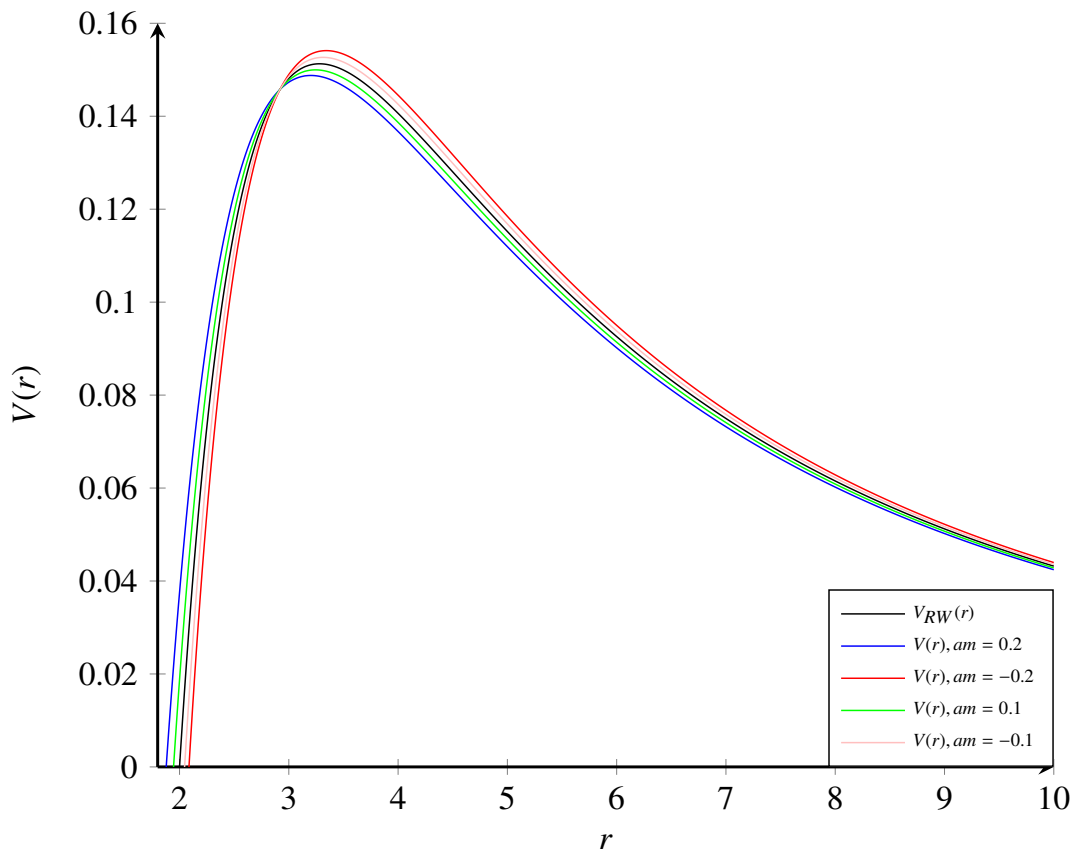

We can also see that around the peak of the potential, $r_0\approx 1.5 R,$ deviations of the potentials on FIG\ref{grafzoom} can be interpreted as Zeeman-like splitting. 
Similar effect was observed in the case of charged scalar field in a vicinity of the Reissner-Nordstr\"{o}m black hole \cite{scalar1, scalar2}.

\begin{figure}[h]
\centering
	\begin{tikzpicture}[scale=1.4]
\node[draw=none] at (1, 2) {$\ell = 2$};
\node[draw=none] at (1, 4.1) {$\ell = 3$};
\node[draw=none] at (1, 7) {$\ell = 4$};
\begin{axis}[
    axis lines = left,
    axis line style = thick,
    xlabel = \(r\),
    ylabel = {\(V(r)\)},
xmin=1.8,
    xmax=14,
    ymin=0.00,
    ymax=0.72,
    legend cell align={left},
    yticklabel style={/pgf/number format/fixed},
    legend style={font=\tiny,at={(1,1)},anchor=north east}]
\addplot [
    domain=1.7:14, 
    samples=500, 
    color=black,
]
{(x-2)*(2*(2+1)*x-6)/x^4};
\addlegendentry{\(V_{RW}(r)\)}

\addplot [
    domain=1.7:14, 
    samples=500, 
    color=blue,
    ]
    {(x-2)*(2*(2+1)*x-6)/x^4+0.2*(2*(2+1)*(6-2*x)*x+2*(5*x-16))/(2*x^5)};
\addlegendentry{\(V(r), am=0.2\)}

\addplot [
    domain=1.7:14, 
    samples=500, 
    color=red,
    ]
    {(x-2)*(2*(2+1)*x-6)/x^4-0.2*(2*(2+1)*(6-2*x)*x+2*(5*x-16))/(2*x^5)};
\addlegendentry{\(V(r), am=-0.2\)}

\addplot [
    domain=1.7:14, 
    samples=500, 
    color=black,
]
{(x-2)*(3*(3+1)*x-6)/x^4};

\addplot [
    domain=1.7:14, 
    samples=500, 
    color=black,
]
{(x-2)*(4*(4+1)*x-6)/x^4};

\addplot [
    domain=1.7:14, 
    samples=500, 
    color=blue,
    ]
    {(x-2)*(3*(3+1)*x-6)/x^4+0.2*(3*(3+1)*(6-2*x)*x+2*(5*x-16))/(2*x^5)};

\addplot [
    domain=1.7:14, 
    samples=500, 
    color=blue,
    ]
    {(x-2)*(4*(4+1)*x-6)/x^4+0.2*(4*(4+1)*(6-2*x)*x+2*(5*x-16))/(2*x^5)};

\addplot [
    domain=1.7:14, 
    samples=500, 
    color=red,
    ]
    {(x-2)*(3*(3+1)*x-6)/x^4-0.2*(3*(3+1)*(6-2*x)*x+2*(5*x-16))/(2*x^5)};

\addplot [
    domain=1.7:14, 
    samples=500, 
    color=red,
    ]
    {(x-2)*(4*(4+1)*x-6)/x^4-0.2*(4*(4+1)*(6-2*x)*x+2*(5*x-16))/(2*x^5)};
\end{axis}
\end{tikzpicture}
\caption{Plot of the potential for $R=2$ ($M=1$) and $\ell=2,3,4$.} \label{grafl}
\end{figure}
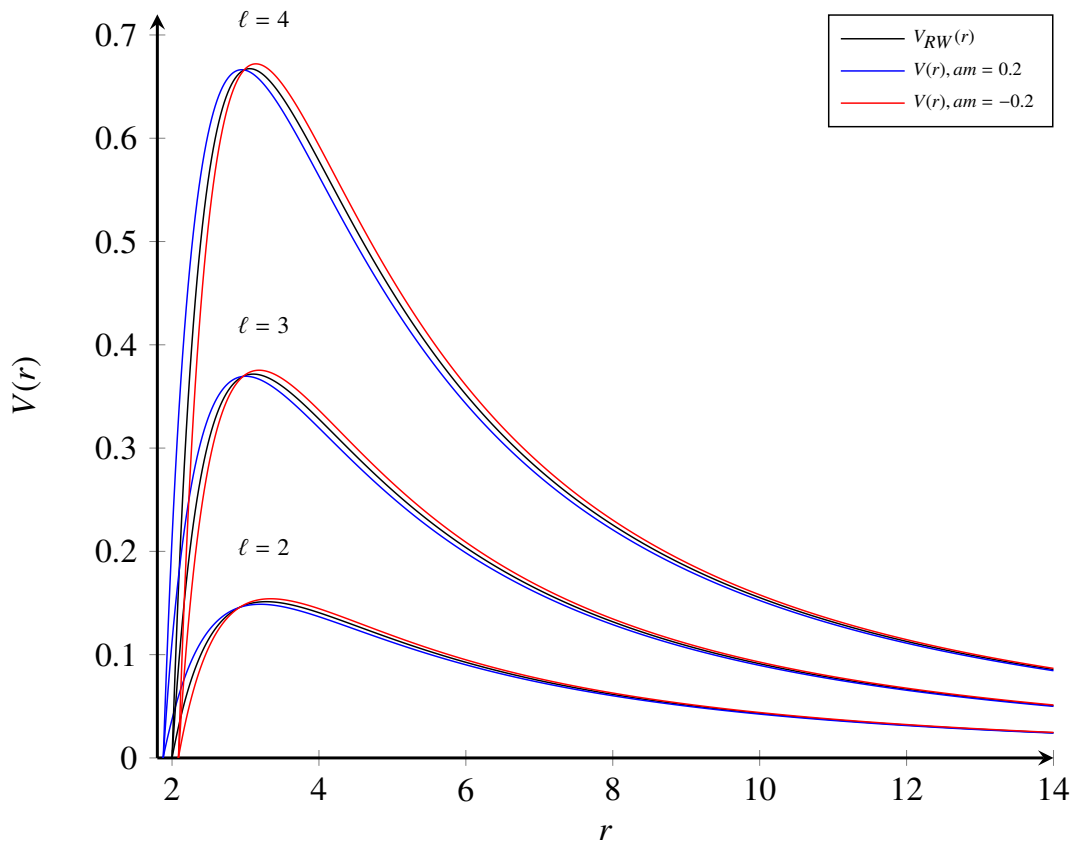

The property of Zeeman-like spliting and a unique peak around $r_0\cong 1.5 R$ persists also for higher multipoles as illustrated in FIG \ref{grafl}.
Calculation of the QNM frequencies for the potential \eqref{master} using the WKB method is given in \cite{letter}. There we see that the NC effects manifest in a significant decrease  (in an absolute value) of the imaginary part of the QNM frequency, which  points to the conclusion that noncommutativity destabilizes the system.

\section{Concluding remarks and discussion}

Utilizing the framework of Hopf algebras, we employed a Drinfeld twist to induce the spacetime noncommutativity.
We then  applied the formalism of twisted noncommutative geometry based on  Hopf algebras with a purpose of formulating  a noncommutative theory of gravity in a bottom-up approach.
As a final outcome, we came out with a proposal for  the noncommutative vacuum Einstein equation.
This proposal turns out to have a correct and smooth commutative limit, it is  devoid of any inconsistencies and after all, it appears to be the only one that  allows for nontrivial solutions.
Its deeper origins  and  a general mathematical structure, including the extension to a divergentless quantity, namely the noncommutative Einstein tensor, will be addressed in a future work.

As a next step,  the proposed noncommutative Einstein equation  was implemented in the context
of linearized gravitational perturbation theory  on an arbitrary background and was shown to give rise
to the equations of motion that are of  the first order in both the perturbation and deformation parameter.
Specifying  this procedure to the Schwarzschild background and for the axial perturbations in particular, 
we obtained the noncommutative corrections to the  Regge-Wheeler potential. 
The axial perturbations of the Schwarzschild black hole in the regime of noncommutative spacetime thus appear to be driven by the quantum corrected   Regge-Wheeler potential
of the form (\ref{master}).
\par

Regarding the obtained quantum corrected potential, a feature that instantly comes on spot
 is the effect of blurring or smearing of the black hole  horizon.
This feature is readily manifested on figures $1-3,$ which show  that different multipoles $(\ell, m)$ of the perturbation interact with different gravitational potentials, especially
in the region close to the horizon. The same conclusion is further confirmed
by inspecting the zeros of the potential, showing a  small $\lambda a$-dependent  shift around the core central value $R$ driven by the spacetime noncommutativity.
We point out that a similar feature reminiscent of the Zeeman-like splitting was already observed in the case of charged scalar field in a vicinity of the Reissner-Nordstr\"{o}m black hole \cite{scalar1, scalar2}.

Having  reduced the gravitational scattering problem to an equivalent problem of scattering in quantum mechanics, where the scattering potential has the form of the corrected Regge-Wheeler barrier (\ref{master}), it would be of  interest to investigate  a rate of energy loss at which gravitational radiation takes the energy away from the compact binary system, which
may consist of two black holes or neutron stars that rotate around each other. This would tell us how discrete structure of spacetime would influence the process of orbital decay. The energy radiated by a compact binary system may be calculated within the framework of black hole
perturbation theory\footnote{Another approach involves the post-Newtonian theory \cite{blanchet,blanchet1,blanchet2, blanchet3, damour}.
However, as this approach is  based upon the slow-motion and 
weak-field approximation and the orbital velocity  during the late stages of the inspiral
 falls rather deep into a relativistic regime, 
the post-Newtonian expansion  does seem inadequate in this situation. Instead, a
required level of accuracy in the final stages of the inspiral is more likely  achieved within the framework of black hole perturbation theory.} \cite{poisson1,poisson2,poisson3,poisson4, poisson5, Poisson:1994yf}.
As a matter of fact, these  orbital decay and  dissipative processes were the first instance where the reality of gravitational waves was confimed, albeit in an indirect way.
Namely, the existence of gravitational waves   had been  indirectly confirmed in the  late  1970s  through the  measurements  of  the  orbital  decay of the  Hulse–Taylor binary  pulsar \cite{hulse-taylor}, on what occasion the observed energy loss due to gravitational radiation in the Hulse–Taylor binary turned out to almost perfectly match the result predicted by the Einstein quadrupole formula.
In this context it is worthy to note that 
the rate of energy loss at which  gravitational waves pull out energy from the system, obtained  along the lines of references \cite{poisson1, poisson2, poisson3, poisson4, poisson5, Poisson:1994yf}, contains the Einstein qudrupole formula as its leading contribution.
 At the present time, pulsars  are  again at  the  forefront  of  the  search  for tracking down gravitational waves, this  time  from  binary  systems of  central   galactic  black  holes.

Indeed, recently reported findings of the North American Nanohertz Observatory for Gravitational Waves (NANOGrav)
point toward the evidence  for a Gravitational-wave Background (GWB), i.e. gravitational waves  with  slowly  evolving   frequencies that contribute  to  a noise-like  broad  band   signal in  the nHz  range \cite{clanak},\cite{NANOGrav:2023gor}. It is believed that they originate from
supermassive black-hole binaries  with  masses of $10^{8} - 10^{10} $ solar masses. NANOGrav findings resulted from
studying the pulsar timing data set  gathered on the basis of continual observation of $67$ pulsars over a period of $15$ years
and   identifying the statistical correlations among them.

Another topic of significant interest  involves the tidal deformability of the compact astrophysical objects.
This feature is encoded within the so called tidal Love numbers (TLNs) \cite{love, relativisticlove, relativisticlove1, relativisticlove2, relativisticlove3} which describe the  response of a rigid body and in particular any self-gravitating object to external tidal forces.
TLNs of the self-gravitating object immersed in a tidal environment depend on the object's internal structure and the properties of the external gravitational field.
Distinctive feature of Einstein's GR is that all tidal Love numbers for black holes identically vanish. Under this circumstance any minute deviation of TLNs from zero, deduced from gravitational-wave measurements in a binary inspiral,
would be a smoking-gun of new physics at the horizon scale. TLNs different from zero would either  signal that the observed object is not a black hole, but some other exotic compact object (if we asuume that the GR is correct) or that the observed object is indeed a black hole, but GR needs an upgrade at the horizon scale and in the strong field regime. 
In this context  TLNs  may  serve as an instrument for discriminating between various proposals for  an accurate  theory of gravity valid at the horizon scale \cite{Cardoso:2017cfl}.

 It is therefore of  significant priority to  determine the exact analytical form of TLNs for quantum corrected Regge-Wheeler potential  that we obtained here. This  potential embodies a particular theory of gravity, the one that goes beyond the standard GR and incorporates in itself a noncommutative nature of spacetime. If TLNs for the Schwarzschild black hole in this particular theory of gravity  turn out to be different from zero, this 
would mark a distinguishing characteristic which might single out this particular proposal among all other proposals 
aiming at realistic description of gravity in the strong field regime and horizon scale conditions.

We plan to extend our analysis in several different directions. Primarily, it  remains to redo the calculations for the polar perturbations in order to get noncommutative/quantum corrections to the Zerilli potential. Moreover, it would be tempting to  delve into higher orders in the noncommutativity parameter and encompass more realistic backgrounds that we are more likely to encounter in the astrophysical environment, such as the Kerr metric. The latter would bring the results of our study closer to the experimentalist forefront.
\par
From the perspective of experimentalists, the primary concern revolves around the spectrum. The QNM spectrum was initially computed in \cite{letter} directly using the WKB method, revealing that the noncommutative effects do not significantly affect the linear stability of the black hole. 
To delve deeper into the analysis of this spectrum, more robust techniques such as the Leaver's continued fraction method are necessary.
\\

\noindent{\bf Acknowledgment}\\
This  research was supported by the Croatian Science
Foundation Project No. IP-2020-02-9614 \textit{Search for Quantum spacetime in Black Hole QNM spectrum and Gamma Ray Bursts}. The authors would like to thank Kumar. S. Gupta for discussions. A.S. and N.H. would like to thank Nikola Konjik for useful comments and discussion. Part of the calculations were checked using the Mathematica package.


\appendix

\section{Perturbations of Schwarzschild black hole in the $a \to 0$ limit} \label{app:rw}
Here we present a brief overview of the calculations for the perturbation of the Schwarzschild black hole in the $a \to 0$ limit. 
This procedure was pioneered by Regge and Wheeler \cite{rw} in 1957. 
Modern and concise review of the subject is presented in \cite{rous}, which will serve as our main reference. Behaviour of the tensor spherical harmonics under parity transformation is elaborated in more detail in \cite{proceeding}.\par
We start by splitting the total metric $g_{\mu \nu}$ into the Schwarzschild background $\oo{g}_{\mu \nu}$ and perturbation $h_{\mu \nu}$:
\begin{equation}
	g_{\mu \nu} = \oo{g}_{\mu \nu} + h_{\mu \nu}.
\end{equation}
Exploiting the spherical symmetry, one can decompose the perturbation into spherical harmonics derived from the standard spherical harmonics
\begin{equation}
	Y_{\ell m}(\theta, \varphi) = \sqrt{\frac{(2 \ell + 1)(\ell - m)!}{4 \pi (\ell + m)!}}P_{\ell m}(\cos \theta)\: e^{i m \varphi}.
\end{equation}
The perturbation metric in spherical basis is
\begin{equation}
\begin{aligned}
	h_{t \theta} & =\frac{1}{\sin \theta} \sum\limits_{\ell, m} h_0^{\ell m}(r)\partial_{\varphi} Y_{\ell m}(\theta, \varphi)e^{-i \omega t}, \quad h_{t \varphi}=-\sin \theta \sum_{\ell, m} h_0^{\ell m}(r) \partial_\theta Y_{\ell m}(\theta, \varphi)e^{-i \omega t} ,  \\
	h_{r \theta} & =\frac{1}{\sin \theta} \sum\limits_{\ell, m} h_1^{\ell m}(r) \partial_{\varphi} Y_{\ell m}(\theta, \varphi)e^{-i \omega t}, \quad h_{r \varphi}=-\sin \theta \sum_{\ell, m} h_1^{\ell m}(r) \partial_\theta Y_{\ell m}(\theta, \varphi)e^{-i \omega t}, \\
	h_{a b} & =\sin \theta \sum\limits_{\ell, m} h_2^{\ell m}(r) \epsilon_{c(a} D^c \partial_{b)} Y_{\ell m}(\theta, \varphi) e^{-i \omega t},
\end{aligned}
\end{equation}
where the indices $a$ and $b$ in the last equation belong to the set $\{\theta, \varphi\}$. $\epsilon_{ab}$ is the totally antisymmetric symbol such that $\epsilon_{\theta \varphi} = +1$ and $D_a$ is the 2-dimensional covariant derivative associated with the metric of the 2-sphere,  $d\theta^2 + \sin^2{\theta} d\varphi^2$. All the other components of the axial perturbations vanish.
\par
Due to the invariance of the theory under space-time diffeomorphisms, the parametrization with the functions $h_0^{\ell m}, h_1^{\ell m}$ and  $h_2^{\ell m}$ is redundant.
Let us consider a small (comparable to $h_{\mu \nu}$) change of coordinates 
\begin{equation} \label{diffeo}
x^\mu \to x^\mu + \xi^\mu. 
\end{equation}
This induces the transformation
\begin{equation} \label{transformedMetric}
	h_{\mu \nu} \to h_{\mu \nu} + \nabla_{\mu} \xi_\nu + \nabla_{\nu} \xi_\mu
\end{equation}
at the linear level. 
Here the covariant derivative and lowering of indices is done using the background metric.
In the axial sector, the nonzero components of the generator $\xi^\mu$ that preserve the odd parity of the perturbations can be decomposed into spherical harmonics as follows:
\begin{equation}
	\xi_\theta=\sum_{\ell, m} \xi^{\ell m}(r) \partial_\theta Y_{\ell, m}(\theta, \varphi)e^{-i \omega t}, \quad \xi_{\varphi}=\sum_{\ell, m} \xi^{\ell m}(r) \partial_{\varphi} Y_{\ell, m}(\theta, \varphi) e^{-i \omega t},
\end{equation}
and the induced gauge transformations on the functions $h_0, h_1$ and $h_2$ are
\begin{equation}
	h_0 \rightarrow h_0 + i \omega \xi, \quad h_1 \rightarrow h_1-\frac{d\xi}{d r}+\frac{2}{r} \xi, \quad h_2 \rightarrow h_2-2 \xi,
\end{equation}
where we dropped the $\ell m$ for simplicity.
As a consequence, one can always choose a gauge in which $h_2^{\ell m} = 0,$ which is the well-known Regge-Wheeler (RW) gauge for the axial perturbations \cite{rous}. Notice that this gauge choice is possible for $\ell \geq 2$ only.\footnote{
For the  monopole ($\ell = 0$) and dipole ($\ell = 1$) case, the function $h_2$ is identically zero and it can be shown that $h_0$ and $h_1$ can be gauged away by $\xi^\mu$, i.e. they are nondynamical. 
The monopole mode of the axial gravitational perturbations represents a change in the black hole mass and the dipole mode corresponds to a shift in the location and value of the black hole's angular momentum.
}
We drop the indices $\ell$ and $m$ in the equations below for clarity, since at the linear level no coupling between modes with different values of these parameters is expected.
\par
In the RW gauge the perturbation is
\begin{equation} \label{rwgauge}
\begin{array}{ll}
	h_{t \theta}=\frac{1}{\sin \theta} \sum\limits_{\ell, m} h_0^{\ell m}(r) \partial_{\varphi} Y_{\ell m}(\theta, \varphi)e^{-i \omega t}, & h_{t \varphi}=-\sin \theta \sum\limits_{\ell, m} h_0^{\ell m}(r) \partial_\theta Y_{\ell m}(\theta, \varphi)e^{-i \omega t}, \\
	h_{r \theta}=\frac{1}{\sin \theta} \sum\limits_{\ell, m} h_1^{\ell m}(r) \partial_{\varphi} Y_{\ell m}(\theta, \varphi) e^{-i \omega t}, & h_{r \varphi}=-\sin \theta \sum\limits_{\ell, m} h_1^{\ell m}(r) \partial_\theta Y_{\ell m}(\theta, \varphi) e^{-i \omega t}.
\end{array}
\end{equation}
Using this ansatz we obtain the following expression for the Ricci tensor:
\begin{equation}
        R_{\mu \nu} =
        \begingroup
\renewcommand*{\arraystretch}{1.6}
\begin{pmatrix}
        0 & 0 & \frac{\dd_\varphi}{\sin \theta}Y_{\ell m}(\theta, \varphi)\frac{- i}{4 r^4} M_1(r) & \sin \theta \: \dd_\theta Y_{\ell m}(\theta, \varphi) M_1(r) \\
        0 & 0 & \frac{\dd_\varphi}{\sin \theta}Y_{\ell m}(\theta, \varphi)\frac{- i}{4 r^4 (r - R)^2} M_2(r) & \sin \theta \dd_\theta Y_{\ell m}(\theta, \varphi) \frac{1}{4 r^4(r - R)^2} M_2(r) \\
        \text{Sym} & \text{Sym} & \frac{(\sin \theta \dd^2_{\theta \varphi} - \cos \theta \dd_\varphi)}{1 - \cos^2 \theta}Y_{\ell m}(\theta, \varphi) \frac{i}{r^2} M_3(r) & (\sin \theta \dd^2_{\theta \theta} - \cos \theta \dd_\theta - \frac{1}{\sin \theta} \dd^2_{\varphi \varphi}) Y_{\ell m}(\theta, \varphi) \frac{-1}{2 r^2} M_3(r) \\
        \text{Sym} & \text{Sym} & \text{Sym} & (\sin \theta  \dd^2_{\theta \varphi} - \cos \theta \dd_\varphi)Y_{\ell m}(\theta, \varphi) \frac{i}{r^2} M_3(r)
\end{pmatrix},
\endgroup
\end{equation}
where
\begin{align}
	M_1(r) &= 2r (2R - \ell(\ell + 1)r) h_0 + 4 i r^2 \omega (r - R) h_1 + 2 r^3(r - R)(i \omega h_1' + h_0''), \label{appricci2} \\
	M_2(r) &= 4 i r^4 (r - R)\omega h_0 + 2 r^2 (r - R) \big(r^3 \omega^2 - (r - R)(\ell(\ell + 1) - 2)\big)h_1 - 2 i \omega r^5(r - R) h_0', \label{appricci1} \\ 
	M_3(r) &= \frac{i r^3 \omega}{r - R}h_0 + R h_1 + r(r - R)h_1'.  \label{appricci3}
\end{align}
The vacuum Einstein equation reduces to three radial equations $M_1 = 0, ~ M_2 =0, ~  M_3 = 0$.
We have three equations, but only two functions $h_0$ and $h_1$. 
It can be shown, however, that these three equations are linearly dependent (see (\ref{dependence1}) in appendix \ref{app:dependence}).
We express $h_0$ from equation (\ref{ricci3}) and plug it into equation (\ref{ricci1}) to get
\begin{equation}
\Big( \big( 2 - \ell(\ell + 1) \big) r^2 + \big(\ell(\ell + 1) - 6\big) R r + 5 R^2 + \omega^2 r^4 \Big) h_1
 + 
r(r - R)\Big( (5R - 2r)h_1' + r(r - R)h_1''\Big) = 0.
\end{equation}
We then substitute $h_1(r) = \frac{r^2}{r - R}Q(r)$ so that the above equation becomes
\begin{equation}
	\frac{(r - R)(3R - \ell (\ell + 1) r) + r^4 \omega^2}{r^2 (r - R)^2}Q + \frac{R}{r(r - R)}Q' + Q'' = 0.
\end{equation}
After introducing the tortoise coordinate $r_*$ satisfying $\frac{d r}{d r_*} = 1 - \frac{R}{r}$, we have
\begin{equation} \label{eqRW}
	\begin{split}
		&\frac{d^2 Q}{d r_*^2} + \Big( \omega^2 - V_{RW}(r) \Big) Q = 0, \\
		&V_{RW}(r) = \frac{(r - R)\big(\ell (\ell + 1)r - 3R\big)}{r^4}.
	\end{split}
\end{equation}
$V_{RW}(r)$ is the Regge-Wheeler potential. 
It vanishes at the horizon and the spatial infinity and is characterized by the single peak near the light ring at $\frac{3}{2}R$.

\section{Dependence of the Ricci components} \label{app:dependence}

To be on equal footing with \cite{rous} in the $a \to 0$ limit, we multiply (\ref{ricci2}) by $\frac{1}{4 r^2}$ and \ref{ricci1}) by $\frac{1}{4 r^2 (r - R)^2}$. Components (\ref{ricci2}), (\ref{ricci1}) and (\ref{ricci3}) become respectively
\begin{align}
\begin{split}
	E_{t \varphi} &= \frac{1}{4 r^2}\bigg[ 2r (2R - \ell(\ell + 1)r) h_0 + 4 i r^2 \omega (r - R) h_1 + 2 r^3(r - R)(i \omega h_1' + h_0'') \\
	&\quad+ \lambda a \Big(  (2\ell(\ell + 1)  r + R)h_0 + i r \omega (4 r - 3 R) h_1 + r (4r - 5 R) h_0' + r^2 R ( i\omega h_1' + h_0'')  \Big)\bigg],
\end{split} \label{appricci11} \\
\begin{split}
	E_{r \varphi} &= \frac{1}{4r^2(r-R)^2}\bigg[\: 4 i r^4 (r - R)\omega h_0 + 2 r^2 (r - R) \big(r^3 \omega^2 - (r - R)(\ell(\ell + 1) - 2)\big)h_1 - 2 i r^5(r - R) h_0' \\
	&\quad+ \lambda a \Big( 2 i r^3 \omega (r - 2R) h_0 + \big(2(\ell(\ell + 1)+12) r (r- R)^2 - 9(r - R)^2 R - r^4 R \omega^2\big)h_1 + i r^4 R \omega h_0' + 2 r (r - R)^3 h_1' \Big) \bigg],
\end{split} \label{appricci22} \\
\begin{split}
	E_{\theta \varphi} &= \frac{i r^3 \omega}{r - R}h_0 + R h_1 + r(r - R)h_1'  \\
    &\quad - \lambda a \Big( \frac{i r^2 R \omega}{2(r - R)^2}h_0 - 3 \frac{r - R}{r} h_1 - \frac{1}{2}R h_1'\Big).
\end{split} \label{appricci33}
\end{align}

From \cite{rous} we infer that components of the Ricci tensor satisfy the dependence relation (\ref{dependence1}) below with $E(r) = 0$.
In our case, however, the same dependence relation produces additional $\lambda a E(r)$ term on the right:
\begin{equation} \label{dependence1}
	\frac{d E_{r \varphi}}{d r} + \frac{i r^2 \omega}{(r - R)^2} E_{t \varphi} + \frac{R}{r(r - R)}E_{r \varphi} + \frac{\ell(\ell + 1) - 2}{2 r(r - R)}E_{\theta \varphi} = \lambda a E(r), 
\end{equation}
where
\begin{equation}
\begin{split}
	E(r) &= i \omega \frac{2\ell(\ell+1)r^2 + (5 - 3\ell(\ell + 1))Rr - R^2}{4(r - R)^3} h_0 + i r \omega \frac{6r^2 - 15 r R + 8R^2}{4(r - R)^3}h_0'+ i \omega \frac{R r^2}{2(r - R)^2} h_0'' \\
	&+ \Big( (r - R)^2 \big( 4 (\ell(\ell + 1) - 6)r^2 - 2(\ell(\ell + 1) - 27)R r - 27R^2 \big) - \omega^2 r^4 (4r^2 - 7R r + 2 R^2)\Big)h_1  \\
	&+ \big(2(\ell(\ell + 1) + 6)r^3 - (3\ell(\ell + 1) + 31) R r^2 + (\ell(\ell + 1) + 24)R^2 r - 5R^3 - 2 R r^4 \omega^2 \big)h_1' + \frac{r - R}{2r}h_1''.
\end{split}
\end{equation}
Therefore we need to generalise this dependence relation by finding $a$-linear differential operators which act on components of the Ricci tensor and produce zero on the right.
\par
Note that in the expressions for the $E_{r \varphi}$ and $E_{\theta \varphi}$ there appear no second derivatives of $h_0$ and $h_1$.
This means that the most general relation is of the form
\begin{align}
	&\frac{d E_{r \varphi}}{d r} + \frac{i r^2 \omega}{(r - R)^2} E_{t \varphi} + \frac{R}{r(r - R)}E_{r \varphi} + \frac{\ell(\ell + 1) - 2}{2r(r - R)}E_{\theta \varphi}  \\
	&+\lambda a \Big( C^0_{r \varphi} E_{r \varphi} + C^1_{r \varphi}\frac{d E_{r \varphi}}{d r} + C^0_{t \varphi}E_{t \varphi} + C^0_{\theta \varphi}E_{\theta \varphi} + C^1_{\theta \varphi}\frac{d E_{\theta \varphi}}{d r} \Big) = 0,
\end{align}
where $C^i_{\mu \nu}(r)$ are some radial functions. This reduces to 
\begin{equation} \label{dependence2}
	C^0_{r \varphi} E_{r \varphi} + C^1_{r \varphi}\frac{d E_{r \varphi}}{d r} + C^0_{t \varphi}E_{t \varphi} + C^0_{\theta \varphi}E_{\theta \varphi} + C^1_{\theta \varphi}\frac{d E_{\theta \varphi}}{d r} - E = 0
\end{equation}
when (\ref{dependence1}) is used.
Each $C^i_{\mu \nu}$ multiplies some linear combination of $(h_0, h_0', h_0'', h_1, h_1', h_1'')$. By extracting coefficients in front of these functions it is possible to write the equation (\ref{dependence2}) in matrix form as
\begin{equation}
	\begin{pmatrix}
	C^0_{r \varphi}\\ C^1_{r \varphi} \\ C^0_{t \varphi} \\ C^0_{\theta \varphi} \\ C^1_{\theta \varphi} \\ -1
	\end{pmatrix}^T
\left(
\begin{array}{cccccc}
	\frac{i r^2 \omega}{r - R} & - \frac{i \omega r^3 }{2(r - R)} & 0 & \frac{1}{2} \Big( 2 -\ell(\ell + 1) + \frac{\omega^2 r^3}{r - R} \Big) & 0 & 0 \\ 
	\frac{i \omega r (r - 2R)}{(r - R)^2} & \frac{i \omega R r^2}{2(r - R)^2} & -\frac{i \omega r^3}{2(r - R)} & \frac{\omega^2 r^2(2r - 3R)}{2(r - R)^2} & \frac{1}{2} \Big(2 -\ell(\ell + 1) + \frac{\omega^2 r^3}{r - R}\Big) & 0 \\ 
	\frac{R}{r} - \frac{1}{2}\ell(\ell + 1) & 0 & \frac{1}{2} r (r - R) & i \omega (r - R) & \frac{1}{2}i \omega r (r - R) & 0 \\
	\frac{i \omega r^3}{r - R} & 0 & 0 & R & r(r - R) & 0 \\
	\frac{i \omega r^2 (2r - 3R)}{(r - R)^2} & \frac{i \omega r^3 }{r - R} & 0 & 0 & 2r & r(r - R) \\
	E_1 & E_2 & E_3 & E_4 & E_5 & E_6 \\
\end{array}
\right)
	\begin{pmatrix}
		h_0 \\ h_0' \\ h_0'' \\ h_1 \\ h_1' \\ h_1''
	\end{pmatrix} = 0,
\end{equation}
where
\begin{align}
	E_1 &=  i \omega \frac{2\ell(\ell + 1)r^2 + (5 - 3\ell(\ell + 1)) r R - R^2}{4(r - R)^3}, \\
	E_2 &=	i \omega r \frac{6 r^2 - 15 r R + 8 R^2}{4(r - R)^3}, \\
	E_3 &=  i \omega \frac{R r^2}{2(r - R)^2}, \\
	E_4 &= \frac{   (r - R)^2 \big( 4 (\ell(\ell + 1) - 6)r^2 - 2(\ell(\ell + 1) - 27)R r - 27 R^2\big) + \omega^2 r^4(-4 r^2 + 7 R r - 2R^2)   }{4 r^3(r - R)^3}, \\
	E_5 &= \frac{2 (\ell(\ell + 1) + 6)r^3 - (3\ell(\ell + 1) + 31)R r^2 + (24 +\ell(\ell + 1))R^2 r - 5 R^3 - 2 \omega^2 R r^4}{4r^2 (r - R)^2}, \\
	E_6 &= \frac{r-R}{2r}.
\end{align}
Existence of such functions $C^i_{\mu \nu}$ is guaranteed by the vanishing determinant of the above matrix. 
This proves that components (\ref{appricci11}) - (\ref{appricci33}) are linearly dependent up to the first order in $a$ and therefore $\hat{R}_{(\mu \nu)} = 0$ admits nontrivial solutions.

\section{Schr\"odinger form of the radial equation of motion} \label{app:schrodinger}
	To interpret the problem in terms of quantum mechanical scattering on a potential barrier it is necessary to transform both the function $h_1(r)$ and independent variable $r$ in the equation \eqref{radial1}:
\begin{equation}\begin{split}
        &r (r-R) \Big( \ell(\ell + 1) r (R-r) + 2r^2 - 6rR + 5R^2 + \omega^2 r^4 \Big) h_1 + r^2(r - R)^2\Big( (5R - 2r)h_1' + r(r - R)h_1''\Big) \\
         &+  \lambda a \bigg[\Big(\ell(\ell + 1)r(r - R)^2 - 6r^3 + \frac{R}{2}(49
r^2 - 64 r R + 26 R^2 - \omega^2 r^4) \Big) h_1 + r(r - R)^2 \Big( 3 ( r- 2R) h_1' + \frac{1}{2}r R h_1''\Big) \bigg]=0.
\end{split} 
\end{equation}
First, we substitute $h_1(r) = \frac{r^2}{r - R}Q(r)$ so that the equation becomes
\begin{equation}
\begin{split}
	&\frac{(r - R)(3R - \ell (\ell + 1) r) + r^4 \omega^2}{r^2 (r - R)^2}Q + \frac{R}{r(r - R)}Q' + Q'' \label{qeq}\\
		&+\lambda a \Big( \frac{2 \ell (\ell + 1) r (r - R)^2 - (3r - 2R)(2r - R)(r - 2R) - r^4 R \omega^2}{2 r^3 (r - R)^3}Q + \frac{(r - 2 R)(3r - 2R)}{r^2(r - R)^2}Q' + \frac{R}{2r(r - R)} Q'' \Big) = 0. 
\end{split}
\end{equation}
With this substitution Regge and Wheeler were able to reduce $a \to 0$ limit of this equation to the Schr\"odinger equation \eqref{eqRW} by introducing the tortoise coordinate $r_*$ satisfying $\frac{d r}{d r_*} = 1 - \frac{R}{r}$.
The same substitution and tortoise coordinate in the full equation (\ref{qeq}) do not work because of the additional $a$-linear terms.
Requiring that in $a \to 0$ limit our equation becomes (\ref{eqRW}) fixes the most general form of substitution and $\star$-tortoise coordinate $\hat{r}_*$:
\begin{equation} \label{subs}
	Q(r) = \Big(1 + \lambda a \: \psi(r)\Big)W(r), \qquad \frac{d r}{d \hat{r}_*} = \Big( 1 - \frac{R}{r} \Big)\Big(1 + \lambda a \: \xi(r) \Big).
\end{equation}
To make the calculations shorter it is benefitial to cast equation (\ref{qeq}) in condensed form as
\begin{equation} \label{ec1}
	\Big( A + \lambda a B + f^{-2}(1 + \lambda a C)\omega^2 \Big)Q + \Big(\frac{R}{r^2 f} + \lambda a D \Big) Q' + \Big(1 + \lambda a E\Big)Q'' = 0,
\end{equation}
where $f(r) = 1 - \frac{R}{r}$ and
\begin{equation} \label{data}
	\begin{split}
		&A(r) = \frac{3R -\ell(\ell + 1)r }{r^3 f}, \qquad B(r) = \frac{2\ell(\ell + 1)r(r - R)^2-(3r - 2R)(2r - R)(r - 2R)}{2 r^6 f^3}, \\
	&C(r) = \frac{-R}{2 r^2 f}, \qquad D(r) = \frac{(r - 2 R)(3 r - 2R)}{r^4 f^2}, \qquad E(r) = \frac{R}{2 r^2 f}.
	\end{split}
\end{equation}
Keep in mind that $'$ indicates derivative with respect to $r$ everywhere $\Big(\frac{d}{d \hat{r}_*}$ is written explicitly$\Big)$. 
Using (\ref{subs}) equation (\ref{ec1}) becomes
\begin{equation} \label{ec2}
	\begin{split}
		&\bigg( A + \lambda a \Big(B + A \psi + \frac{R}{r^2 f} \psi' + \psi'' \Big) + \omega^2 f^{-2} \Big( 1 + \lambda a (C + \psi) \Big) \bigg) W  \\
		&+ \lambda a \bigg( D + 2 \psi' - \xi' - \frac{R}{r^2 f}E \bigg) W' + f^{-2}\bigg( 1 + \lambda a (E - 2 \xi + \psi) \bigg) \frac{d^2 W}{d \hat{r}_*^2} = 0.
	\end{split}
\end{equation}
Requiring that this equation after multiplication by some radial function\footnote{Function is allowed to have $a$-linear terms.} becomes of the form
\begin{equation} \label{Vnc}
	\frac{d^2 W}{d \hat{r}_*^2} + \Big( \omega^2 - \big(V_{RW}(r) + a V_{NC}(r)\big) \Big) W = 0, \\
\end{equation}
i.e. that the noncommutativity is present in the effective potential only, is equivalent to requiring
\begin{align}
	D + 2 \psi' - \xi' - \frac{R}{r^2f}E &= 0, \label{req1} \\
	C + \psi = E - 2 \xi + \psi&. \label{req2}
\end{align}
Equation (\ref{req1}) ensures there is no $W'$ term in (\ref{ec2}) and equation (\ref{req2}) makes sure that terms multiplying $\omega^2$ and $\frac{d^2 W}{d \hat{r}_*^2}$ can be set to $1$ after dividing the whole equation (\ref{req2}) with $f^{-2}\Big( 1 + \lambda a (E - 2 \xi + \psi) \Big) = f^{-2}\Big( 1 + \lambda a (C + \psi)\Big)$.
Condition (\ref{req2}) gives solution for $\xi = \frac{1}{2}(E - C)$ and by plugging this into (\ref{req1}) we get $\psi$. Explicitly,
\begin{equation} \label{psixi}
	\xi = \frac{R}{2 r(r - R)}, \qquad \psi = \frac{1}{2}\Big(\frac{3}{r} - \frac{1}{r - R} + \frac{1}{R} \log \frac{r}{r - R} \Big).
\end{equation}
After dividing  equation (\ref{ec2}) with $f^{-2}\Big( 1 + \lambda a (E - 2 \xi + \psi) \Big) = f^{-2}\Big( 1 + \lambda a (C + \psi)\Big)$ we get
\begin{equation}
	\bigg( f^2 A + \lambda a f^2 \Big( B + \frac{R}{r^2 f} \psi' + \psi'' - A C) \Big) + \omega^2 \bigg) W + \frac{d^2 W}{d \hat{r}_*^2} = 0.
\end{equation}
To get (\ref{master}) one just needs to substitute the expressions from (\ref{data}) and (\ref{psixi}).\\

\end{document}